\documentclass[journal]{vgtc}                     

\onlineid{1760}

\vgtccategory{Analytics and Decisions}

\title{FZ-VIS: A Visual Analytics Framework for Quantities-of-Interest-Aware Scientific Lossy Compression}

\author{
  \authororcid{Guoxi Liu}{0000-0002-8164-7185},
  \authororcid{Yuxiao Li}{0000-0002-8715-5982}, 
  \authororcid{Congrong Ren}{0009-0006-6285-7271}, 
  \authororcid{Robert Underwood}{0000-0002-1464-729X}, 
  \authororcid{Xin Liang}{0000-0002-0630-1600}, 
  \authororcid{Bei Wang}{0000-0002-9240-0700},\texorpdfstring{\\}{ }
  \authororcid{Sheng Di}{0000-0002-7339-5256}, 
  \authororcid{Franck Cappello}{0000-0002-7890-3934}, and 
  \authororcid{Hanqi Guo}{0000-0001-7776-1834}
}

\authorfooter{
  \item
  	Guoxi Liu, Yuxiao Li, Congrong Ren, and Hanqi Guo are with The Ohio State University.
  	E-mail: \{liu.12722 | li.14025 | ren.452 | guo.2154\}@osu.edu.
  \item 
  	Robert Underwood, Sheng Di, and Franck Cappello are with Argonne National Laboratory.
  	E-mail: \{runderwood | sdi1 | cappello\}@anl.gov.
  \item 
    Xin Liang is with Oregon State University. E-mail: lianxin@oregonstate.edu.
  \item 
    Bei Wang is with University of Utah. E-mail: beiwang@sci.utah.edu.
}

\abstract{Modern scientific simulations generate massive volumes of data, making lossy compression essential for efficient storage and transmission. However, preserving critical quantities of interest (QoIs) under lossy compression is inherently data- and task-dependent, requiring domain scientists to navigate complex trade-offs between compression ratio and data fidelity. Exploring these trade-offs often involves large design and evaluation spaces, motivating human-in-the-loop approaches that combine interactive exploration with quantitative analysis.
To address this challenge, we present FZ-VIS, an interactive framework for human-in-the-loop \textbf{\textit{feature}}-oriented lossy \textbf{\textit{compression}} design and \textbf{\textit{visual}} analytics. FZ-VIS provides a web-based interface for rapidly generating and comparing compression configurations, along with integrated visualization tools for assessing reconstruction fidelity and QoI preservation through both visual inspection and quantitative metrics. We demonstrate the utility of FZ-VIS through case studies involving three representative user groups: novice users selecting compression methods, compressor developers examining internal pipeline behavior, and domain scientists investigating feature preservation. The case studies show how FZ-VIS helps users efficiently navigate complex design spaces and make informed decisions that balance compression performance with application-specific QoI requirements.}


\teaser{
  \centering
  \includegraphics[width=0.9\linewidth, alt={Overview of FZ-VIS showing a workflow composition panel, linked quantitative and spatial comparison views, and QoI-specific topology and spectrum analysis modules.}]{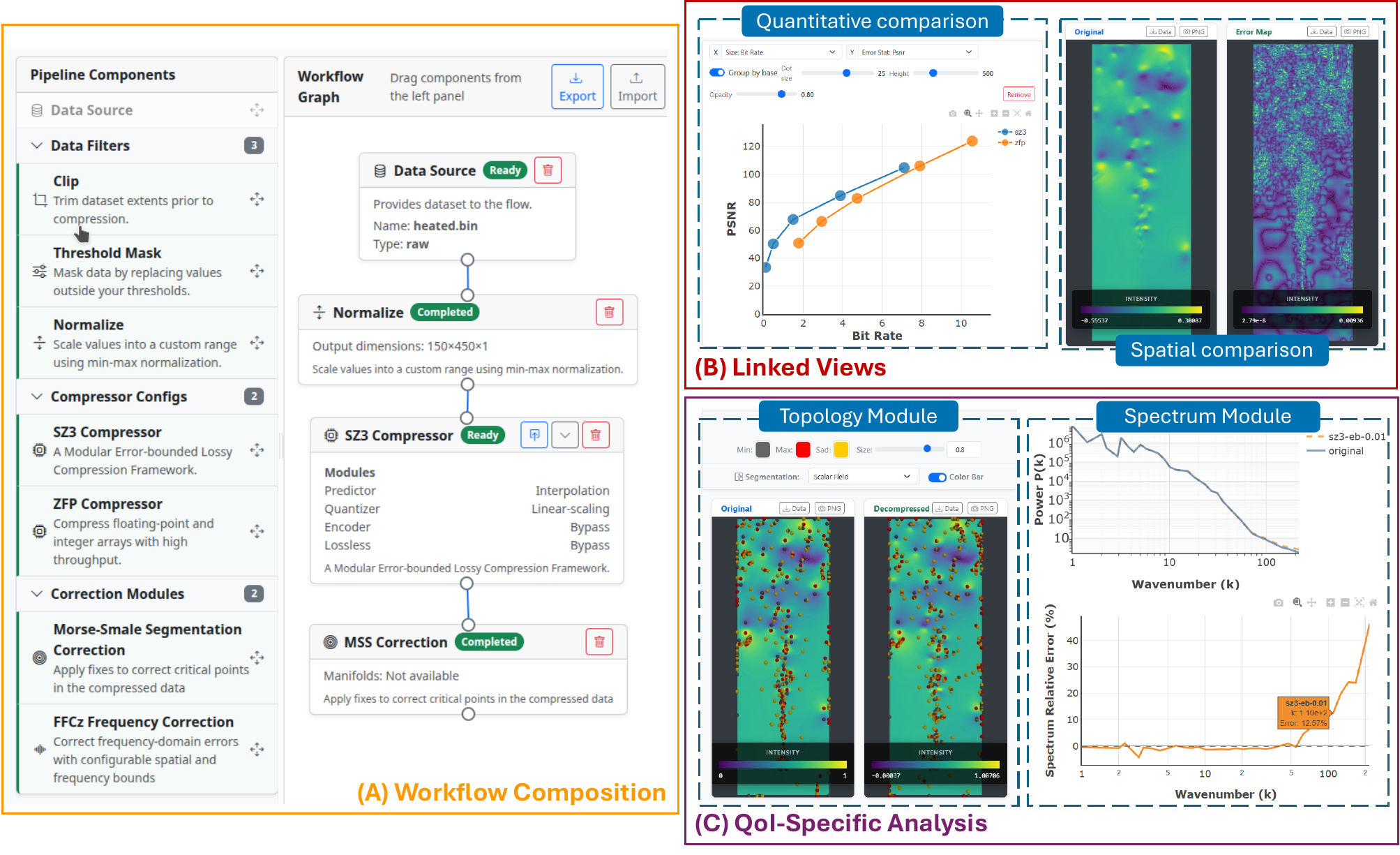}
  \caption{%
  	FZ-VIS supports QoI-aware lossy compression through (A) workflow composition, (B) linked quantitative and spatial views, and (C) QoI-specific analysis for topology- and spectrum-related fidelity validation.
  }
  \label{fig:teaser}
}

\graphicspath{{figs/}} 

\usepackage{amsmath}
\usepackage{amsfonts}
\usepackage{tabu}                      
\usepackage{booktabs}                  

\usepackage{mathptmx}                  
\usepackage{wrapfig}
\usepackage{url}

\begin{document}

\firstsection{Introduction}
\maketitle

Modern scientific workflows can generate vast amounts of data when running large-scale parallel applications or performing simulation tasks. In many situations, the volume of data produced is too large to efficiently transmit over networks, store in storage systems, or analyze using standard user tools. For example, the state-of-the-art Energy Exascale Earth System Model (E3SM)~\cite{DOE2023E3SM}, funded by the U.S. Department of Energy, provides a kilometer-scale land model to accurately model geographical characteristics and extreme weather occurrences, which can generate terabytes of data from just a single-year simulation. Such large data sizes pose challenges in the visualization, comprehension, and prediction of climate patterns.

To address the challenge of data storage, transmission, and analysis, lossy compressors are considered a practical solution for drastically reducing the data size for scientific research, while still carefully controlling the loss of data accuracy based on user demand. However, scientific datasets often consist of floating-point data arrays, which lossless compressors such as GZIP~\cite{Deutsch1996Gzip} cannot compress effectively~\cite{Baker2016Evaluating}.
Therefore, many lossy compressors targeting floating-point data arrays have been developed, some of which are considered ideal solutions for climate simulations by guaranteeing the validity of data after decompression~\cite{Baker2016Evaluating,Sasaki2015Exploration}. Furthermore, many error-bounded lossy compression techniques have been proposed to reduce data size while controlling the distortion~\cite{Liang2018ErrorControlled,Zhao2020Significantly,Zhao2021Optimizing,Zhao2022MDZ,Li2021Resilient,Liu2021Exploring,Yu2022Ultrafast,Lindstrom2014FixedRate,Liang2023SZ3}. However, many other scientific research applications also require lossy compression, and the performance of existing lossy compressors on their data remains unclear. Moreover, users need the ability to select the compressors best suited to their specific requirements.

In practice, domain scientists often seek to preserve specific quantities of interest (QoIs) when applying lossy compression. For example, in the analysis of high-energy diffraction microscopy (HEDM) data, peaks associated with grain orientations and strain states in crystalline materials are essential for grain identification and the interpretation of internal strain distributions~\cite{Xia2024Preserving,Li2026pMSz}. Contour trees are critical for supporting a variety of post hoc visualization tasks in applications ranging from materials science to climate simulation~\cite{Zhou2009Automatic,Kopp2023Temporal,Gorski2025General}. Similarly, the power spectrum of cosmological data is a key QoI for analyzing the distribution of matter and energy across spatial scales. These users therefore need a consistent, flexible, and high-performance interface to use and understand the effects of compression on these quantities. LibPressio~\cite{Underwood2021Productive} was developed to provide such a general and uniform user interface for compressor users, simplifying existing compressor workflows and reducing the large design space for applying different compressors. However, it lacks a general graphical user interface for users to easily configure compressors and visually evaluate compressor performance on their data.

Popular visualization tools like ParaView~\cite{Ayachit2015ParaView}, Python Matplotlib~\cite{Hunter2007Matplotlib}, and R ggplot2~\cite{Wickham2016ggplot2} are widely used for direct and post-processing scientific data visualization. However, they lack native support for lossy compressor composition and comprehensive metric comparisons. While tools such as Z-checker~\cite{Tao2019Z-checker} and Foresight~\cite{Grosset2020Foresight} address visual evaluation of lossy compressor performance, they overlook QoI evaluation and preservation in scientific applications. To address this gap and streamline QoI-driven lossy compression workflows, we introduce a novel framework that provides an intuitive interface for users to rapidly configure and evaluate hundreds of lossy compressor settings, visually and quantitatively compare their compression quality and performance, and make informed decisions about QoI preservation.

Our main contributions are as follows:
\begin{itemize}[leftmargin=*, noitemsep]
    \item We formulate QoI-aware lossy compression design as a visual analytics problem that tightly couples the \emph{compressor design space} with the \emph{scientific evaluation space}. This formulation highlights why existing compressor interfaces and evaluation tools are inadequate for supporting QoI-driven compression workflows. 
    \item We introduce FZ-VIS, a \textbf{feature}-oriented lossy \textbf{compression} design and \textbf{visual} analytics framework that supports interactive compressor composition, systematic exploration of configuration series, and linked analysis across quantitative metrics, spatial comparisons, and intermediate compressor outputs within a unified environment.
    \item We develop QoI-aware analysis workflows for topology- and spectrum-related tasks, allowing users to assess the impact of lossy compression on topological structures and power spectra while reasoning about trade-offs between compression performance and downstream scientific fidelity.
    \item Through case studies involving novice users, compressor developers, and domain scientists, we demonstrate that FZ-VIS effectively supports diverse decision-making tasks, including compressor selection, pipeline diagnosis, and QoI-aware refinement under application-specific constraints.
\end{itemize}

\section{Related Work}
\label{sec:related}

In this section, we review related work from four perspectives: error-bounded lossy compression; compressor design and visual analytics tools; topology- and power-spectrum-preserving compression; and ensemble visualization and parameter-space exploration.

\subsection{Error-Bounded Lossy Compression}

Lossy compression has become an essential technique for managing the vast volume of scientific data. Traditional lossless compression methods, such as GZIP~\cite{Deutsch1996Gzip} and Zstandard~\cite{Zstandard}, often fall short in achieving the necessary compression ratios, leading to increased storage and transmission costs. As a result, researchers have turned to lossy compression methods, which sacrifice some degree of fidelity for significantly improved compression performance. Error-bounded lossy compressors can significantly reduce data size while allowing bounded pointwise error between the original and decompressed data within a user-specified error bound, and we refer the reader to~\cite{Di2025Survey} for a comprehensive survey.

Error-bounded lossy compressors can be broadly categorized into two types: \textit{transform-based} and \textit{prediction-based}. Transform-based compressors first transform the data into a different domain, such as wavelet transforms and tensor decomposition, and then compress the transformed representation. For example, ZFP~\cite{Lindstrom2014FixedRate} uses a custom orthogonal block transform to decorrelate data within blocks, producing sparsely distributed coefficients that can be efficiently encoded. SPERR~\cite{Li2023SPERR} is a transform-based compressor that employs the CDF 9/7 discrete wavelet transform. MGARD~\cite{Gong2023MGARD} is another transform-based compressor that uses wavelet-based techniques to compress data on uniform and non-uniform structured grids, as well as unstructured grids.

In contrast, prediction-based compressors exploit data correlations by predicting values from neighboring samples and encoding the resulting prediction errors. The SZ family of compressors~\cite{Di2016Fast, Di2018Optimization, Liang2018ErrorControlled, Tao2017Significantly, Zhao2020Significantly, Zhao2021Optimizing, Liang2023SZ3} exemplifies this approach. For example, SZ1.4~\cite{Tao2017Significantly} uses the Lorenzo predictor combined with linear-scaling quantization to convert prediction residuals into integers, which are then encoded with customized Huffman coding and lossless compressors. SZ3~\cite{Liang2023SZ3} is a modular, composable compression framework that allows for easy creation and customization of various prediction-based lossy compressors. FPZIP~\cite{Lindstrom2006Fast} allows a specified number of bit planes to be ignored, making the data distortion controllable on demand. AE-SZ~\cite{Liu2021Exploring} and SRNN-SZ~\cite{Liu2023Scientific} incorporate neural networks to enhance prediction accuracy and compression performance.

\subsection{Compression with Feature Preservation}

In addition to general-purpose lossy compression techniques, there has been growing interest in developing methods that explicitly preserve important data features. In this work, we focus specifically on features in the topological and frequency domains.

Topological features, such as critical points, persistence diagrams, and Morse-Smale (MS) segmentations, are crucial for understanding the underlying structure of scientific data. Several studies have developed lossy compression techniques that explicitly preserve these features, ensuring that essential topological characteristics are retained after compression. For scalar fields, existing studies have primarily focused on preserving critical points, contour and merge trees. 
LOCP~\cite{Fallin2026Fast} preserves the relative ordering of neighboring data values, referred to as local order, thereby preserving all critical points while achieving high compression ratios. Yan et al.~\cite{Yan2024TopoSZ} introduced TopoSZ, which extends the SZ 1.4 compression algorithm by incorporating topological constraints derived from contour-tree-induced segmentations. Gorski et al.~\cite{Gorski2025General} proposed a general framework that augments existing lossy compressors with topology-preserving capabilities by quantifying the adjustments required to preserve contour trees and encoding these corrections using a custom variable-precision scheme. Soler et al.~\cite{Soler2018Topologically} developed a topology-controlled compression method that adaptively quantizes data within individual topological features to preserve persistence diagrams under a given persistence simplification threshold. More recently, Xia et al.~\cite{Xia2025TspSZ} designed TspSZ to preserve the topological skeleton by extending existing critical-point-preserving compressors to also retain separatrices. MSz~\cite{Li2025MSz, Li2026pMSz} represents one of the first efforts to preserve MS segmentations under error-bounded lossy compression using an edit-based paradigm.
For vector fields, prior work has primarily focused on preserving critical points during compression. For example, Liang et al.~\cite{Liang2020Toward,Liang2023Toward} developed a methodology for preserving critical points in piecewise linear and bilinear vector fields. For tensor fields, TFZ~\cite{Gorski2026TFZ} is a topology-preserving compression framework for 2D second-order tensor fields that preserves degenerate points in symmetric fields and eigenvector and eigenvalue graphs in asymmetric fields during lossy compression.

In addition to topology-driven analysis, many scientific applications rely on frequency-domain representations to capture key features. One particularly important Fourier-derived quantity is the \textit{power spectrum}, which describes the power distribution across different frequency components~\cite{vanDaalen2011Effects}. The shape of the power spectrum curve, characterized by the wavenumber $k$ and power spectrum $P(k)$, provides valuable insight into the underlying structure of the data. For example, peaks in the spectrum indicate dominant periodic components, whereas flat regions suggest white noise with uniformly distributed power. Despite its importance, preserving frequency-domain characteristics during lossy compression remains relatively underexplored. Jin et al. performed fine-grained rate-quality modeling to configure existing compressors~\cite{Jin2021Adaptive}, and the more recent FFCz~\cite{Ren2026FFCz} introduced a novel algorithm explicitly designed to preserve Fourier-domain fidelity, including the power spectrum accuracy.

\subsection{Ensemble and Parameter Space Exploration}

Ensemble visualization is a well-established approach for comparing collections of simulation outputs and analyzing their variability. Early work, such as Ensemble-Vis~\cite{Potter2009Ensemble}, introduced coordinated summary views and linked visual representations to facilitate ensemble analysis. Subsequent studies extended these concepts to uncertainty-aware workflows and multidimensional ensemble analysis~\cite{Chen2015Uncertainty,Wang2019Visualization}, emphasizing the importance of statistical summaries and interactive comparison mechanisms.

Parameter space exploration studies how input parameters affect output behavior, quality, and performance. Result-driven and conceptual frameworks emphasize sensitivity analysis, partitioning, optimization, and trade-off reasoning~\cite{Bruckner2010ResultDriven,Sedlmair2014Visual}. Recent surveys underline that temporal/spatial outputs amplify these needs and may require compact abstractions for high-dimensional parameter spaces~\cite{Piccolotto2023Visual}.

Several systems further bridge ensemble visualization and parameter space exploration. Schneider et al. integrated data space and model space in ensemble learning using visual analytics, enabling users to reason jointly about parameter settings, model families, and resulting outputs~\cite{Schneider2021Integrating}. In scientific simulation, He et al.\ introduced InSituNet to support post hoc parameter space exploration of ensemble simulations through learned image synthesis, demonstrating the importance of scalable surrogates and interactive visual feedback when exhaustive inspection is infeasible~\cite{He2020InSituNet}.

\subsection{Design and Visual Analytics Tools for Compression}

Many lossy compressors have been designed specifically for scientific data. However, their diverse user interfaces and numerous configuration parameters pose significant challenges for users seeking to effectively utilize and compare these compression techniques. To unify the interface and streamline the adoption of these lossy compressors, LibPressio~\cite{Underwood2021Productive} provides a standardized and modular API that allows users to seamlessly integrate, configure, and evaluate multiple compression algorithms within a single framework, thereby drastically reducing the complexity of compression workflow management.

For compressor performance evaluation and analysis, several visualization tools have been proposed to support the design and assessment of lossy compressors. Z-checker~\cite{Tao2019Z-checker} offers a standardized framework for assessing the compression quality and performance of various lossy compressors, enabling detailed analysis of error metrics and data fidelity. Foresight~\cite{Grosset2020Foresight}, in contrast, emphasizes extensibility and reproducibility by facilitating the integration and benchmarking of multiple compressors within customizable evaluation pipelines. 

Despite these advances, existing systems still leave a gap for QoI-aware compression analysis. LibPressio standardizes access to compressors but mainly serves as an API rather than an interactive visual environment for exploring design alternatives and their effects on downstream tasks. Z-checker and Foresight support comparative evaluation, yet focus largely on aggregate compression metrics and conventional fidelity measures. In contrast, our work treats the lossy compression workflow as a visual analytics problem, enabling users to compose configuration families, inspect intermediate compressor behavior, and understand how standard metrics relate to application-specific QoIs.

\section{Background}
\label{sec:background}

This section introduces two domain-specific QoIs used to evaluate the effects of lossy compression: topological structure and frequency-domain fidelity. We first summarize Morse-Smale segmentation, and then review Fourier-based spectral analysis and the power spectrum.

\subsection{Morse-Smale Segmentation}
\begin{wrapfigure}{l}{0.5\linewidth}
    \centering
    \includegraphics[width=\linewidth, alt={Illustration of a Morse-Smale segmentation with colored regions separated by integral lines and red and blue spheres marking local maxima and minima.}]{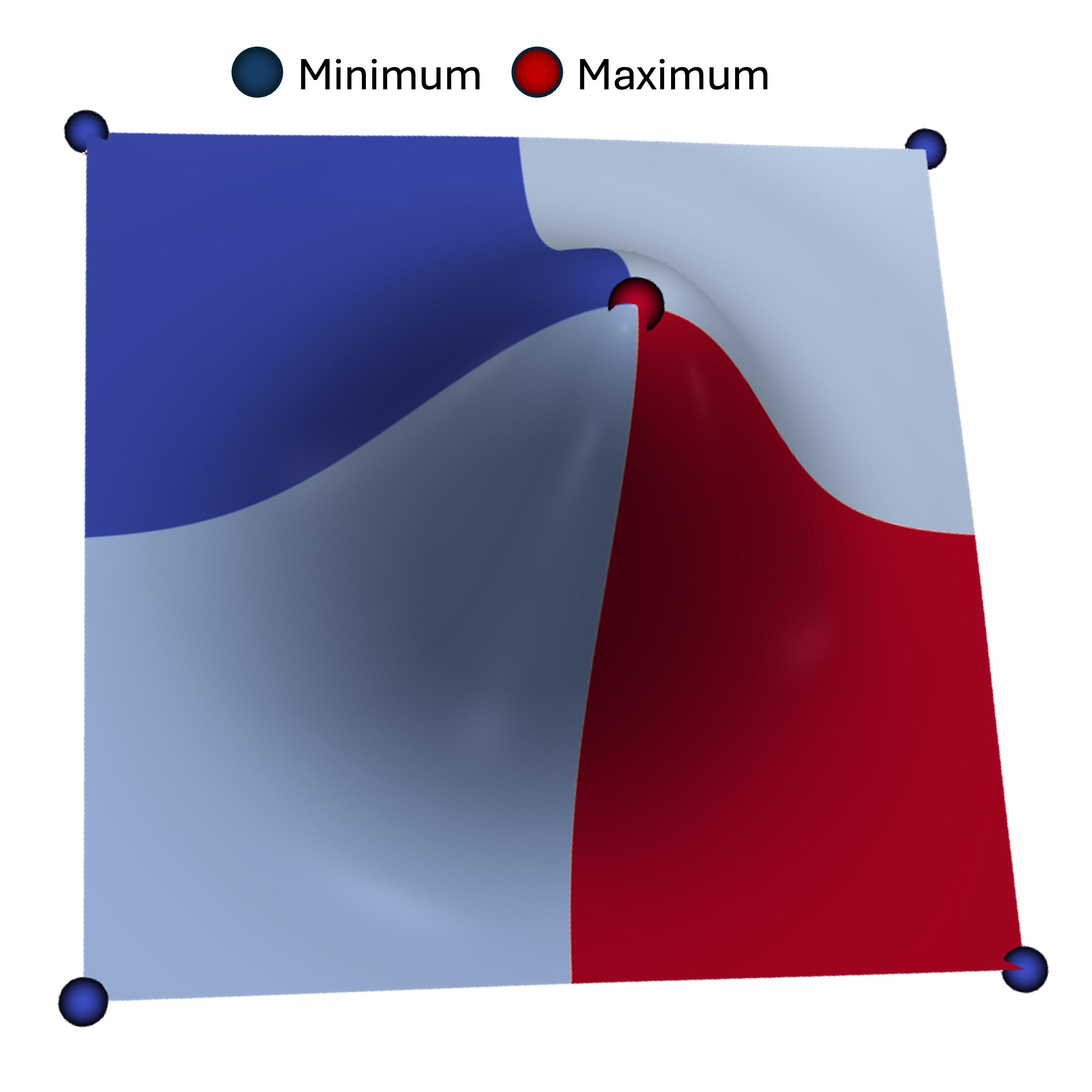}
    \caption{Example of Morse-Smale segmentation. The red and blue spheres represent local maxima and minima, respectively.}
    \label{fig:mss-example}
\end{wrapfigure}

We provide a concise overview of piecewise linear Morse-Smale segmentations (PLMSS)~\cite{Maack2024Parallel}. As depicted in \Cref{fig:mss-example}, for a piecewise linear function defined on a constant triangular or tetrahedral mesh, where each data point $i$ has a unique value $f_i$, PLMSS partitions the data into distinct regions of consistent gradient behavior: all points within the same region follow the steepest ascent and descent paths to the same maximum $M_k$ and minimum $m_j$, respectively.

\emph{Extrema} are data points whose scalar values are higher or lower than those of all their neighbors, depicted as red spheres for maxima and blue spheres for minima in \Cref{fig:mss-example}. \emph{Integral lines} are monotonic paths along mesh edges that follow the steepest ascent or descent direction from a given point until reaching an extremum. \emph{Ascending} and \emph{descending segmentations} group data points that converge to the same minimum and maximum, respectively. PLMSS combines these ascending and descending segmentations to partition the domain into segments. Each segment is defined by a pair of extrema $\langle m_j, M_k \rangle$ and is separated from neighboring segments by integral lines, as illustrated by the differently colored regions in \Cref{fig:mss-example}.

MSz~\cite{Li2025MSz} preserves PLMSS in error-bounded lossy compression by iteratively correcting topological distortions in decompressed data. It takes as input a 2D or 3D scalar field $f$ and its decompressed version $\hat{f}$ obtained from an arbitrary error-bounded lossy compressor with a global error bound $\xi$. The algorithm outputs a set of edits $\delta = \{\delta_i\}$ applied to $\hat{f}$, yielding a corrected field $g = \hat{f} + \delta$ that preserves the PLMSS while strictly satisfying the global error bound $|f_i - g_i| \leq \xi$ for all data points. The core idea of MSz is an edit-based strategy that identifies a subset of points in the decompressed data whose values can be modified within the user-specified error bound to correct topological distortions. More specifically, the method iteratively detects topological distortions, identifies the data point $i$ responsible for each distortion, and applies an edit $\delta_i$ to restore topological consistency. 

\subsection{Fourier and Spectral Analysis}

Techniques based on Fourier transforms are widely used for feature identification and pattern detection. The \emph{Discrete Fourier Transform} (DFT)~\cite{Sundararajan2001Discrete} converts a discrete signal from the spatial domain to the frequency domain. For a 1D dataset with $N$ data points, denoted by $\{x_n\}^{N-1}_{n=0}$, DFT computes its frequency components $X_k$ as follows: 
\[X_k = \sum_{n=0}^{N-1} x_n e^{-i \frac{2 \pi k n}{N}}, k \in \{0, 1, \ldots, N-1\}.\]
DFTs for higher dimensions adhere to a similar pattern, incorporating additional exponential terms for each dimension.

The \emph{power spectrum}, denoted as $P(k)$, illustrates the allocation of power among various frequency bands. For a discrete 3D dataset with values $\{x_{m, n, p}\}$, we first normalize the variations by removing the mean background as follows:
\[x'_{m, n, p} = \frac{x_{m, n, p} - \bar{x}}{\bar{x}},\] 
where $\bar{x}$ is the average value across $\{x_{m, n, p}\}$. This ensures the power spectrum reflects relative structure and variability rather than absolute magnitude. We then shift the zero-frequency component of $\{x'_{m, n, p}\}$ to the center for symmetry between positive and negative frequencies~\cite{NumpyFFTshift}. The spatial data is then mapped to the frequency domain: $X'_{u, v, w} = FFT(x'_{m, n, p})$. The power spectrum accumulates the squared magnitudes of frequencies at equivalent distances from the center:
\[P(k) = \sum_{u^2+v^2+w^2=k^2} |X'_{u, v, w}|^2,\] where $|\cdot|$ indicates the magnitude of a complex number.

FFCz~\cite{Ren2026FFCz} preserves both spatial and frequency-domain accuracy by applying a post hoc correction to decompressed data under user-defined error bounds in both domains. It represents the error of each frequency-component as a linear combination of spatial errors, reformulating the dual-domain constraints as two intersecting hypercubes. Starting from the spatial error vector produced by an off-the-shelf compressor, FFCz iteratively projects the vector onto the two hypercubes until it converges within their intersection and satisfies both constraints. Because FFCz supports independent tolerances for individual spatial and frequency components, it can preserve the power spectrum by applying specific pointwise relative error bounds to the frequency domain.

\section{Requirement Analysis}
\label{sec:requirements}

\begin{figure}[htb]
    \centering
    \includegraphics[width=\linewidth, alt={Diagram mapping three user groups, compressor developers, novice users, and QoI-driven domain scientists, to five design requirements for FZ-VIS.}]{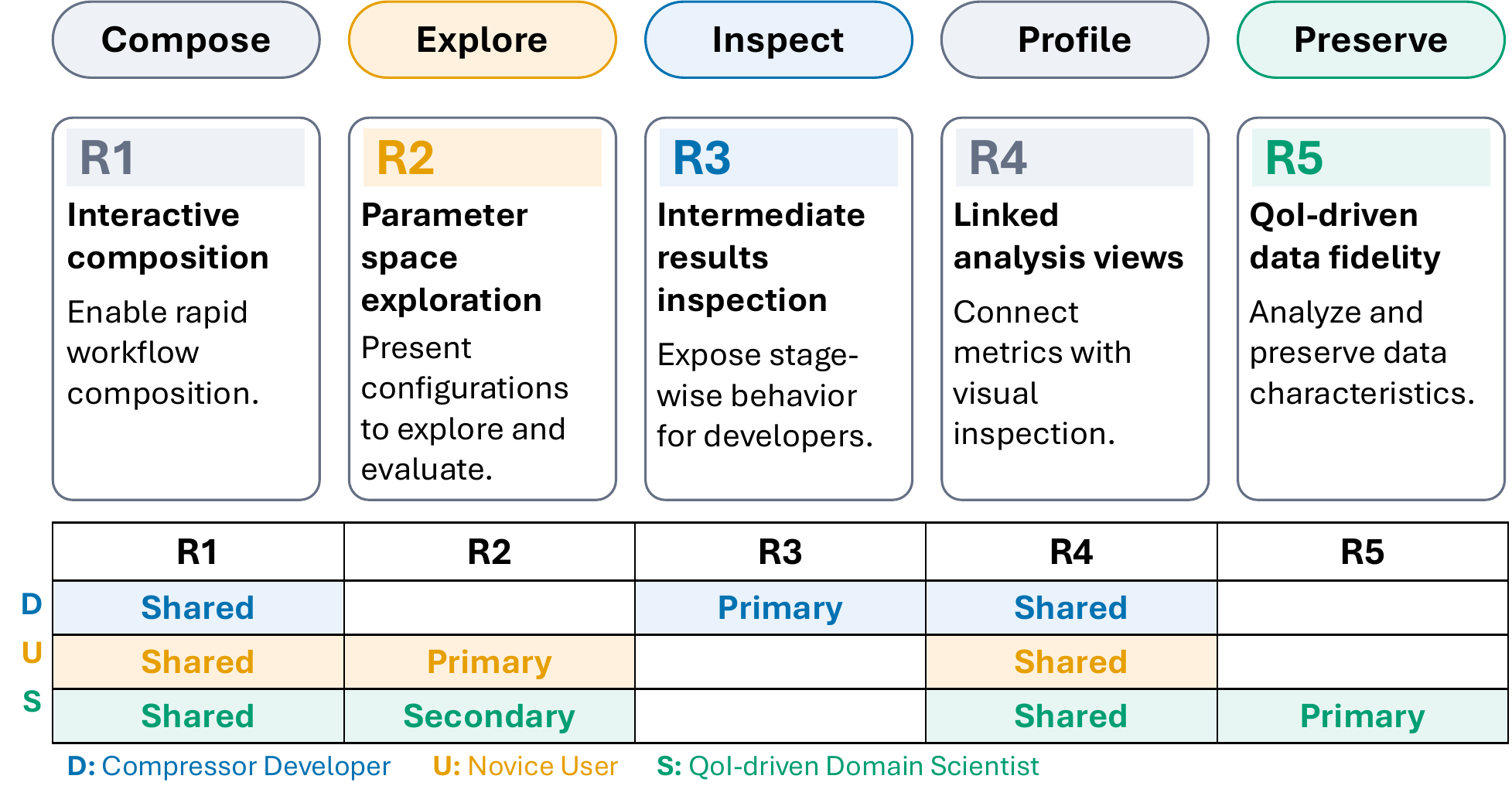}
    \caption{Design requirements derived from three user groups: compressor developers, novice users, and QoI-driven domain scientists.}
    \label{fig:requirements-analysis}
\end{figure}

We derived the following requirements from formative discussions with colleagues who use or study lossy compression in different roles. These discussions helped us identify three primary user groups shown in \Cref{fig:requirements-analysis}: compressor developers (\textit{D}), novice users (\textit{U}), and QoI-driven domain scientists (\textit{S}). Compressor developers focus on creating, debugging, and tuning compressors. Novice users, such as trainees or students, often learn to apply compressors effectively, but may have limited knowledge of compressor internals. QoI-driven domain scientists care less about the compressor implementation itself and more about whether compression preserves the scientific structures, features, or derived quantities needed by downstream analysis.

\textbf{R1 (\textit{D, U, S}) --- Interactively composing compressor workflows.}
Lossy compressors expose many interchangeable modules and parameter choices, and even small configuration changes can substantially impact compression ratio, throughput, and reconstruction quality. All three user groups therefore need an intuitive interface for composing compressor workflows, inspecting available parameters, and modifying configurations without repeated code changes. Such support shortens the iteration cycle for developers, lowers the barrier for novice users, and allows domain scientists to quickly test candidate configurations against their analysis needs.

\textbf{R2 (\textit{U, S}) --- Exploring parameter spaces of compressors.}
This requirement is primary for novice users and also relevant to QoI-driven domain scientists. Novice users, such as trainees and students, often enter the workflow with practical learning goals, such as understanding how compressor parameters affect compression ratio, throughput, and reconstruction quality, but they may not yet know which parameters matter most or how different settings interact. QoI-driven domain scientists also need to explore parameter spaces when searching for configurations that preserve application-specific features under practical performance constraints. Therefore, the system should expose the parameter space in an interpretable manner, helping users move from educational or analytical goals to a manageable set of candidate configurations. Rather than requiring them to tune parameters blindly, it should reveal promising regions of the design space, highlight sensitive parameters and trade-offs, and support iterative exploration from coarse comparisons to focused refinement.

\textbf{R3 (\textit{D}) --- Inspecting intermediate results for debugging and optimization.}
Compressor developers need to understand not only the final compression metrics, but also how each stage of the compressor pipeline behaves internally. They therefore require a system that can expose intermediate results, such as predictions, quantization errors, residuals, encoded outputs, or stage-wise statistics, so that failures can be localized to specific modules or parameter choices. This requirement helps developers diagnose unexpected behavior, compare alternative module designs, and reason about how local changes in the pipeline propagate to the final compression performance.

\textbf{R4 (\textit{D, U, S}) --- Profiling compressor behavior through linked analysis views.}
All user groups need to evaluate the trade-offs induced by lossy compression, but they inspect them from different perspectives: developers study algorithmic behavior, novice users compare configurations to balance performance and quality, and domain scientists verify whether the reconstructed data remains valid for downstream analysis. To support these needs, the system should provide coordinated views of standard performance and fidelity metrics together with direct visual comparison between original and decompressed data. Linking these analysis views helps users move from high-level aggregate summaries to detailed inspection when selecting or rejecting configurations.

\textbf{R5 (\textit{S}) --- Analyzing and preserving QoI-driven fidelity.}
Pointwise error metrics alone are insufficient when downstream tasks depend on preserving specific quantities of interest, such as extrema, Morse-Smale segmentations, or spectral characteristics. QoI-driven domain scientists need analysis tools that expose how compression affects these domain-specific quantities, allow them to compare candidate configurations under QoI-aware metrics, and support region- or feature-specific inspection when failures occur. More importantly, the system should assist users in reasoning about the trade-off between conventional compression objectives and QoI preservation so they can select compressors and settings that satisfy the demands of their specific scientific tasks.

\section{System Design}
\label{sec:system-design}

This section describes how FZ-VIS addresses the requirements mentioned in \Cref{sec:requirements}. FZ-VIS is designed as a human-in-the-loop framework for QoI-aware lossy compression rather than as a collection of isolated interface components. In practice, users do not simply choose a compressor to run once and inspect the final compression ratio and execution time. Instead, they iteratively define an evaluation context, construct candidate configurations, execute batches of experiments, inspect quantitative and visual results, and then refine the search space based on what they learn. FZ-VIS is designed around this iterative loop so that the compressor design space and the evaluation space remain tightly connected throughout the workflow.

At a high level, FZ-VIS serves three complementary user groups. For novice users, it exposes the parameter space of compressors through a guided user interface and supports comparison among candidate configurations that satisfy target performance or fidelity requirements. For compressor developers, it provides a structured workspace for pipeline composition, variant management, and inspection of intermediate results. For domain scientists focusing on QoI in scientific applications, it integrates different feature analysis modules into the compression workflow so that QoI preservation becomes a proactive decision instead of an external post hoc validation step.

\subsection{System Overview}

\begin{figure}[htb]
    \centering
    \includegraphics[width=\linewidth, alt={Architecture diagram of FZ-VIS organized into setup, composition and execution, and analysis and refinement stages connected by an iterative workflow.}]{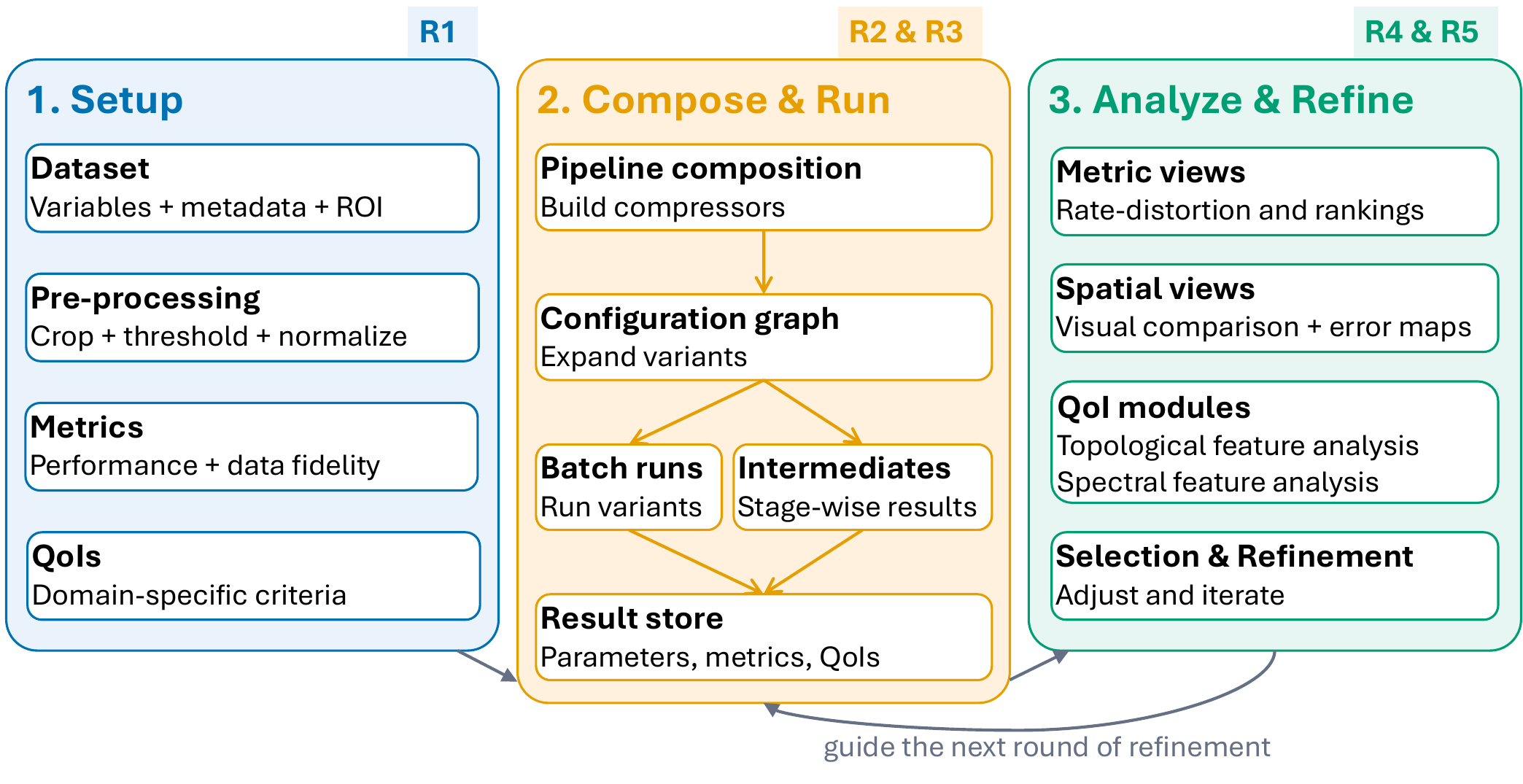}
    \caption{Workflow-oriented architecture of FZ-VIS. The workflow consists of three stages: setup, composition and execution, and analysis and refinement. Each stage produces structured outputs that serve as inputs to the next, while analysis results inform subsequent rounds of configuration refinement.}
    \label{fig:system-design}\end{figure}

\Cref{fig:system-design} summarizes the architecture of FZ-VIS. The system is organized into three major stages: \emph{setup}, \emph{composition and execution}, and \emph{analysis and refinement}.

\begin{itemize}
    \item The \emph{setup} stage establishes the experimental context by defining datasets (e.g., climate simulation data, X-ray CT images, cosmology datasets) with necessary pre-processing operations, standard compression metrics (e.g., bit rate, PSNR, DSSIM), and key QoIs (e.g., local maxima, Morse-Smale segmentations, power spectra). This ensures that experiments are grounded in specific scientific and performance requirements.
    \item In the \emph{composition and execution} stage, users construct and configure compressor pipelines (e.g., SZ3, ZFP) through a guided interface, exploring module options and parameter combinations. This includes specifying different error bounds and managing large families of configurations. Experiments are then executed, recording not only final results but also essential intermediate artifacts for debugging and understanding compressor behavior.
    \item The \emph{analysis and refinement} stage presents linked quantitative views (e.g., rate-distortion curves, frequency error plots) and visual views (e.g., error maps, distributions of critical points, Morse-Smale segmentation boundaries). These views enable users to compare, filter, and refine the design space, making informed decisions on compressor selection to meet specific application needs. This adaptive feedback loop ensures a systematic and iterative refinement of compressor configurations, directly supporting scientifically grounded decisions.
\end{itemize}

This workflow-centered organization is important because it bridges two spaces that are often disconnected in existing tools. On one side is the \emph{compressor design space}, which consists of module choices, parameter values, and execution variants. On the other side is the \emph{scientific evaluation space}, which consists of performance metrics, visual fidelity, and downstream QoIs. In conventional workflows, users often move between command-line scripts, benchmark tables, plotting tools, and domain-specific analysis codes. FZ-VIS unifies these activities in a single environment so that decisions made in one space can be immediately interpreted in the other.

\subsection{Design Space Setup}

FZ-VIS frames compression tuning as a structured design-space exploration task rather than as a sequence of disconnected trial-and-error runs. At the beginning of a study, users define an experiment specification that fixes the dataset, required pre-processing, optional region-of-interest (ROI), evaluation metrics, and any QoI analysis to be applied. This specification establishes a stable analytical context in which many related compressor variants can be generated and compared without repeatedly reconstructing the surrounding workflow.

The purpose of this setup stage is to make the exploration process explicit and reproducible. Instead of executing one compressor configuration at a time, users start from a default pipeline and branch into related alternatives by varying modules, parameters, or shared settings. The resulting variants remain part of the same study definition, so each configuration is interpreted as one point in a larger design space rather than as an isolated benchmark result.

\begin{figure}[htb]
    \centering
    \includegraphics[width=\linewidth, alt={Example FZ-VIS workflow graph paired with a generated JSON specification that records datasets, metrics, QoI modules, and compressor variants.}]{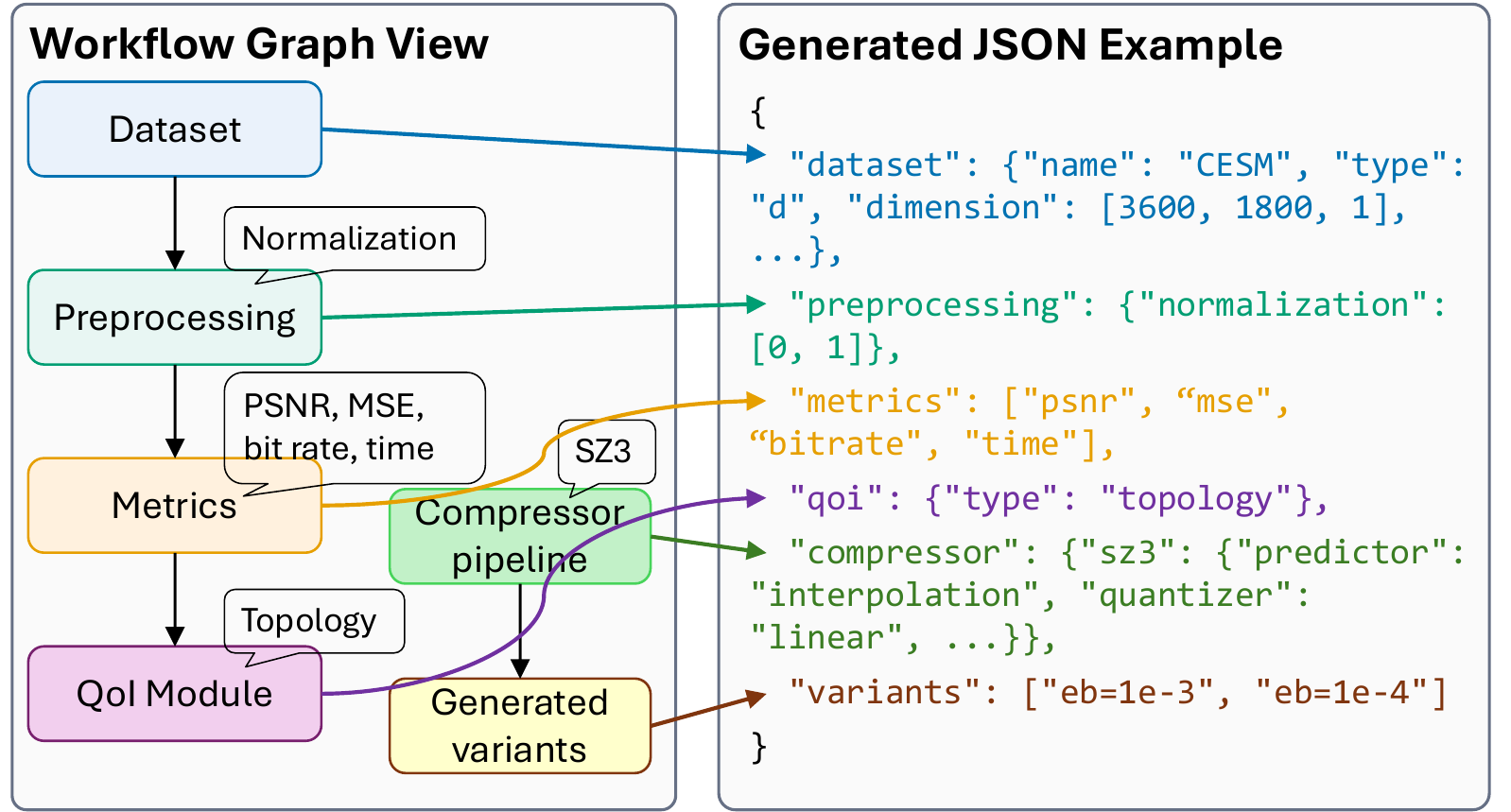}
    \caption{Workflow graph and generated JSON specification in FZ-VIS. The workflow graph defines the experiment structure, including data preparation, evaluation metrics, QoI modules, and compressor variants, while the generated JSON records the same context as an experiment specification. The correspondence between them makes the interactive workflow design reproducible.}
    \label{fig:workflow-spec}
\end{figure}

FZ-VIS also preserves provenance for every explored configuration, including its dataset, pipeline structure, parameter values, execution outputs, derived metrics, and QoI summaries. This provenance is critical for iterative refinement. As illustrated in \Cref{fig:workflow-spec}, the workflow graph is translated into a generated JSON specification that records the same case study definition, enhancing the reproducibility and traceability of the experiment.

\subsection{Compressor Configuration Generation}

A core component of FZ-VIS is the composition of compressor pipelines. It exposes the available modules and parameters of a compressor through a unified interface provided by LibPressio~\cite{Underwood2021Productive}, allowing users to configure internal pipeline modules, such as the predictor, quantizer, encoder, and lossless stage in the SZ3 compressor, without directly editing code or scripts. From the perspective of \textbf{R1}, this interface shortens the iteration cycle by turning low-level compressor configuration into direct interactive manipulation.

\begin{figure}[htb]
    \centering
    \includegraphics[width=\linewidth, alt={Configuration generation interface showing candidate compressor module options presented in parallel coordinates and expanded into a force-directed graph of configurations.}]{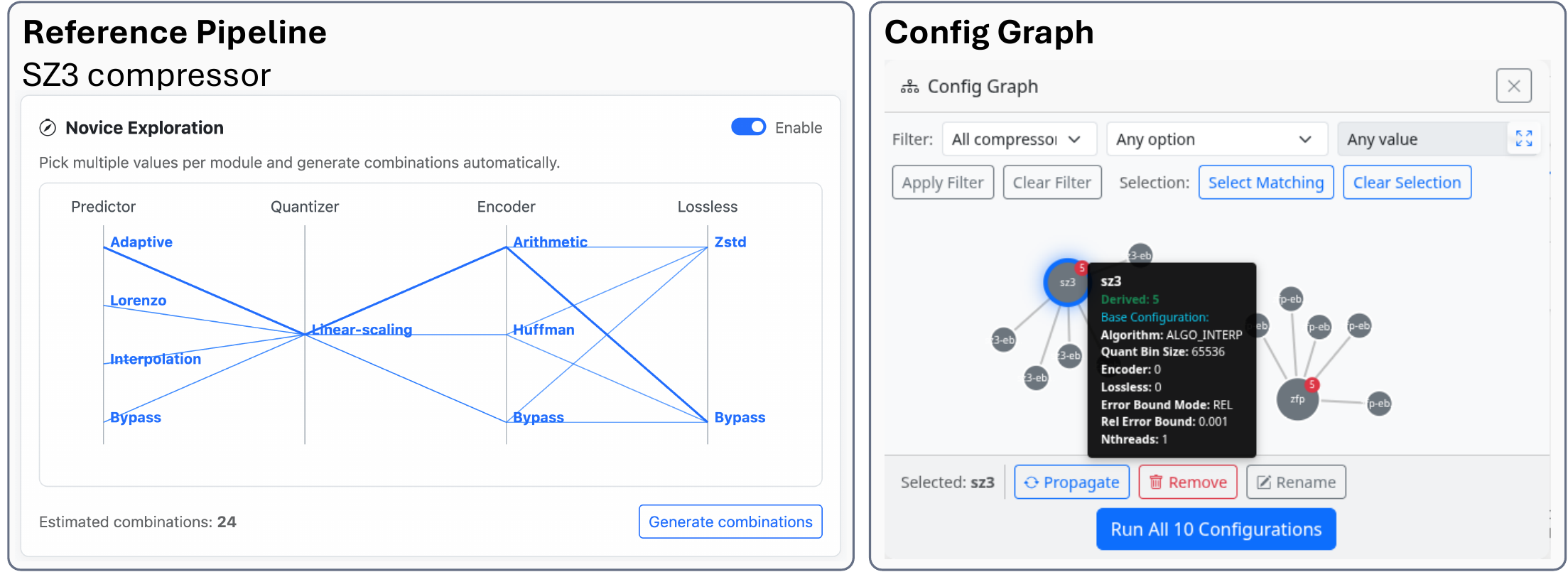}
    \caption{Compressor configuration generation in FZ-VIS. Starting from a reference compressor pipeline, users can combine candidate module options using parallel coordinates. The resulting configurations are organized as a force-directed graph, where each parent configuration can further expand into sub-nodes by propagating parameter values or common execution options.}
    \label{fig:config-generation}
\end{figure}

FZ-VIS is designed to support both single-configuration composition and large-scale configuration generation. A user may begin with a reference compressor pipeline and then derive multiple alternatives by varying modules and parameter ranges. As illustrated in \Cref{fig:config-generation}, the reference pipeline acts as a starting point, while parallel coordinates summarize the candidate choices for different modules. The generated configurations are then organized as parent nodes in a force-directed graph. Each parent configuration can further expand into sub-nodes by propagating parameter values (e.g., error bound) or common execution options (e.g., single-thread, multi-thread, or GPU-accelerated).

This design is particularly important for novice users addressed by \textbf{R2}. Such users may have desired compression requirements, e.g., a minimum compression ratio, a throughput threshold, or a target quality measure, but they often do not know which parameter combinations are worth testing. FZ-VIS therefore does not force them to tune one parameter at a time in isolation. Instead, it presents the parameter space in a form that can be explored through branching, propagation, and side-by-side comparison. The resulting configuration space is explicit and inspectable, making it easier for users to search for candidate settings that match their requirements.

\subsection{Intermediate Inspection and Execution Management}

Once a set of configurations has been defined, FZ-VIS executes the corresponding compression pipelines and records the resulting outputs. The execution layer manages the mapping from a logical compressor configuration to a concrete experiment instance, including the input dataset and the metrics that need to be collected. This stage transforms interactive design decisions into executable experiments that can be compared consistently.

In addition to final compression statistics, FZ-VIS also supports the capture of intermediate results and stage-wise summaries. This capability is essential to \textbf{R3}. For compressor developers, final metrics alone are often insufficient to explain why two related configurations behave differently. Taking SZ3 as an example, a change in the predictor, quantization behavior or entropy characteristics may propagate through the compressor in ways that are not visible from the final compression metrics alone.

This stage therefore serves both experimental execution and debugging. Developers can inspect whether changes in the compressor structure alter residual distributions, whether a stage becomes a performance bottleneck, or whether a configuration behaves anomalously compared with other variants. In this sense, FZ-VIS is not just a benchmark dashboard; it is also a debugging and optimization platform for compressor development.

\subsection{Linked Quantitative and Spatial Analysis}

After experiments are executed, FZ-VIS presents the results through linked quantitative and spatial comparison views. The quantitative comparison view enables users to compare candidate configurations across user-selected metrics, such as bit rate, PSNR, compression ratio, and runtime. It summarizes the tested configurations using plots such as rate-distortion diagrams, trade-off curves, and ranked candidate lists. Users can choose which quantitative measures to compare so that the same result set can support different tasks. For novice users, these views help identify configurations that satisfy target requirements. For developers, they reveal algorithmic trends across compressor variants and expose the trade-offs among fidelity, rate, and execution cost.

These quantitative summaries are tightly coupled with direct visual inspection of the data. FZ-VIS supports side-by-side comparison between original and decompressed data, error-map visualization, and ROI-focused inspection. Linking these visual and quantitative views directly addresses \textbf{R4}. 

\begin{figure}[htb]
    \centering
    \includegraphics[width=\linewidth, alt={Linked analysis interface showing quantitative comparison views for selecting candidate configurations alongside spatial views for comparing original, reconstructed, error-map, and ROI-focused data.}]{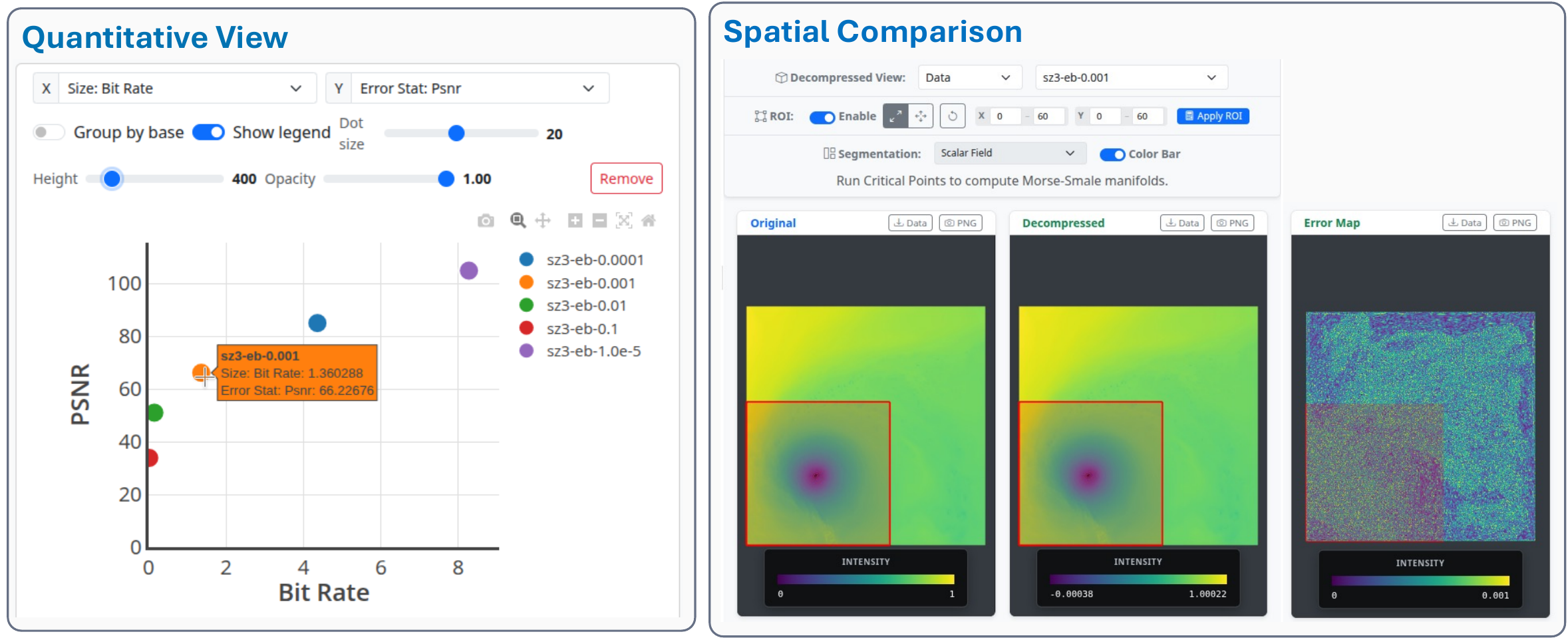}
    \caption{Linked quantitative comparison and spatial analysis in FZ-VIS. The quantitative view on the left shows the metric plot with customizable x- and y-axes, allowing users to select promising configurations from the tested design space. The spatial comparison view on the right shows the original data, the reconstructed data, its error map, and ROI-focused comparison, supporting localized inspection and iterative refinement of the candidate set.}
    \label{fig:linked-analysis}
\end{figure}

As illustrated in \Cref{fig:linked-analysis}, a user can first select a candidate in the quantitative view, inspect its corresponding original, decompressed, and error-map views, and then compare ROI-focused details across multiple candidates before returning to the configuration space to refine the search. This coordination between high-level summaries and localized visual comparisons is essential because numerical measures alone may fail to reveal spatially concentrated artifacts, while visual inspection alone does not scale to large configuration spaces.

Another important aspect of the linked analysis design is selective comparison. Rather than treating all configurations equally at all times, FZ-VIS lets users focus on subsets of configurations that are relevant to the user requirements. For example, a user may examine a region in the quantitative plot, and then compare only those candidates in the ROI-focused view. This filtering-and-linking mechanism progressively narrows the design space and helps users move from coarse exploration to targeted decision-making.

\subsection{QoI-Specific Analysis Modules}

To support domain scientists, FZ-VIS extends beyond generic compression metrics and includes dedicated QoI-specific modules. In the current design, three complementary capabilities are emphasized: topological analysis, spectral analysis, and QoI correction. These modules are integrated into the same environment as the standard performance and fidelity views, so users do not need to export results into separate tools before determining whether a compressor is acceptable for downstream analysis.

The topological analysis module evaluates structures such as critical points and Morse-Smale segmentations. It enables users to compare how different compressor configurations change topological structures, such as the number of false critical points introduced, and which settings preserve the relevant topological structures. This is especially important in applications where downstream analysis depends more on structural correctness than on global pointwise error.

The spectral analysis module supports applications in which the frequency-domain representation is a primary scientific quantity. By computing and comparing power spectra and related spectral relative errors, FZ-VIS allows users to inspect how compression affects frequency-domain representations in addition to spatial fidelity. Users can therefore examine whether a configuration that appears acceptable under spatial error bounds still introduces unacceptable distortion in the frequency domain.

Beyond analysis, FZ-VIS also provides QoI correction modules to assist users when a candidate configuration fails to preserve the target QoI. Instead of treating QoI evaluation as the end of the workflow, these modules use the detected discrepancy to suggest corrective actions, such as applying MSz~\cite{Li2025MSz} to fix Morse-Smale segmentation and FFCz~\cite{Ren2026FFCz} to retain the power spectrum. These modules therefore extend FZ-VIS from an evaluation system into an iterative refinement framework, enabling users to retain desirable candidate configurations by applying QoI-specific corrections.

Above all, these QoI-oriented modules address \textbf{R5} by making QoI preservation an active part of the decision process for domain scientists. Rather than selecting a compressor based only on compression metrics and then validating QoI preservation afterward, users can evaluate candidate configurations under QoI-aware criteria, inspect where and why QoI failures occur, and apply QoI-guided corrections. This unified design supports scientifically grounded compressor selection and enables FZ-VIS to be extended with additional application-specific QoI modules in the future.

\section{Case Study}
\label{sec:case-study}

The key advantage of FZ-VIS is that its linked views guide users through compression design. Rather than assuming that users already know which compressors, parameters, or QoI criteria to prioritize, the system supports an iterative workflow: generating candidates, comparing quantitative and spatial evidence, inspecting failures, and refining the search space. In this section, we demonstrate how FZ-VIS satisfies different requirements through several case studies. These case studies are intended to illustrate integrated workflow support, while achieving the same sequence of configuration, comparison, QoI analysis, and refinement would otherwise require substantially more manual scripting and movement across separate tools.

\subsection{Compressor Selection for Novice Users}
Novice users, such as trainees and students who are new to lossy compression, may be unfamiliar with compressor parameter spaces and how these parameters affect performance. However, they often still have concrete goals for compression behavior, such as target compression ratio and throughput. In this subsection, we use a climate application as a case study to demonstrate how FZ-VIS supports these users in selecting appropriate compressor configurations.

\paragraph{Requirements.} For trainees and students who are new to lossy compression, an important early task is to understand how compressor choices relate to application-specific goals. We illustrate this learning process through a climate application. Climate simulation models have grown increasingly complex as computing resources have advanced, producing far more data than scientists can afford to store in full. As a result, climate users often need lossy compression solutions that substantially reduce storage cost while preserving data quality for downstream analysis. In this case study, we use the CLDHGH field from the CESM~\cite{Hurrell2013Community} dataset. A key quality metric for this application is the Data Structural Similarity Index Measure (DSSIM)~\cite{Baker2024Structural}, and a practical goal is to achieve a compression ratio roughly $2$--$3\times$ higher than lossless compression~\cite{Cappello2025State}. These requirements define the decision context for novice-oriented compressor selection.

\begin{figure}[htb]
    \centering
    \includegraphics[width=\linewidth, alt={CESM case-study comparison showing the original field, a rate-distortion scatter plot for SZ3 predictor variants and ZFP, and error maps for decompressed outputs at one percent relative error.}]{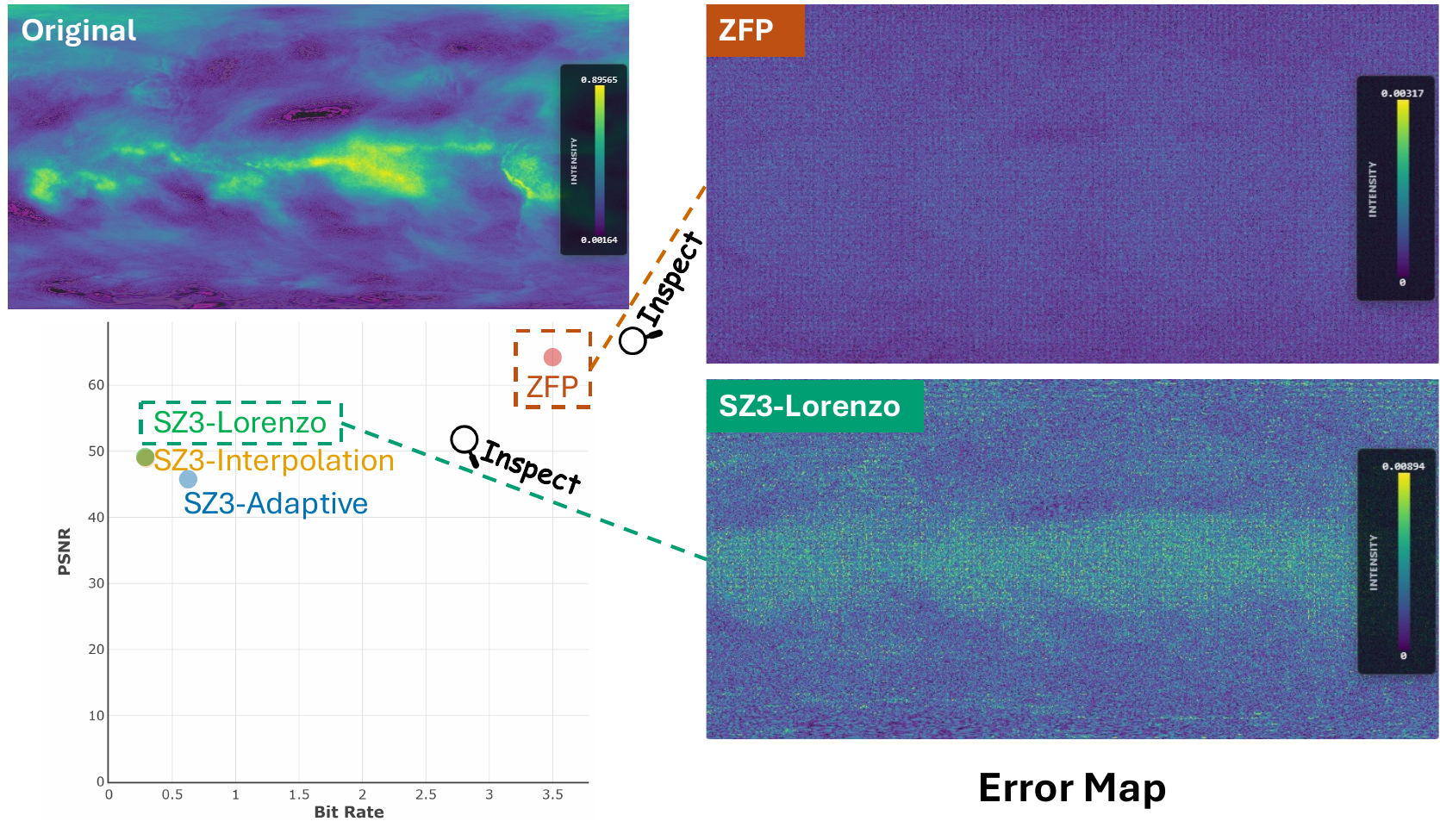}
    \caption{Comparison of SZ3 using different predictors and ZFP on the CESM dataset. Left: the original field visualization and rate-distortion scatter plot of the tested compressor configurations. Right: error maps of decompressed data from SZ3 and ZFP at a $1\%$ relative error bound.}
    \label{fig:cesm-compare-predictor}
\end{figure}

\paragraph{Design Phase.} A novice user typically does not know which compressor family or predictor to try first. FZ-VIS addresses this uncertainty by exposing candidate SZ3 module choices and ZFP settings through a guided composition interface, producing a small starting set instead of many scripted runs. The user begins with three SZ3 predictors (adaptive, Lorenzo, and interpolation-based predictors) and compares them with a ZFP baseline under the same default $1\%$ relative error bound. The linked rate-distortion and spatial views in \Cref{fig:cesm-compare-predictor} show not only that the SZ3 variants achieve lower bit rates while ZFP attains higher PSNR, but also how these differences appear spatially in the reconstructed field. This first comparison gives the user a basis for deciding which configurations merit broader exploration.

\begin{figure}[htb]
    \centering
    \includegraphics[width=\linewidth, alt={CESM metric comparison views plotting SZ3 and ZFP configurations across error bounds to compare bit rate, PSNR, DSSIM, and compression ratio trade-offs.}]{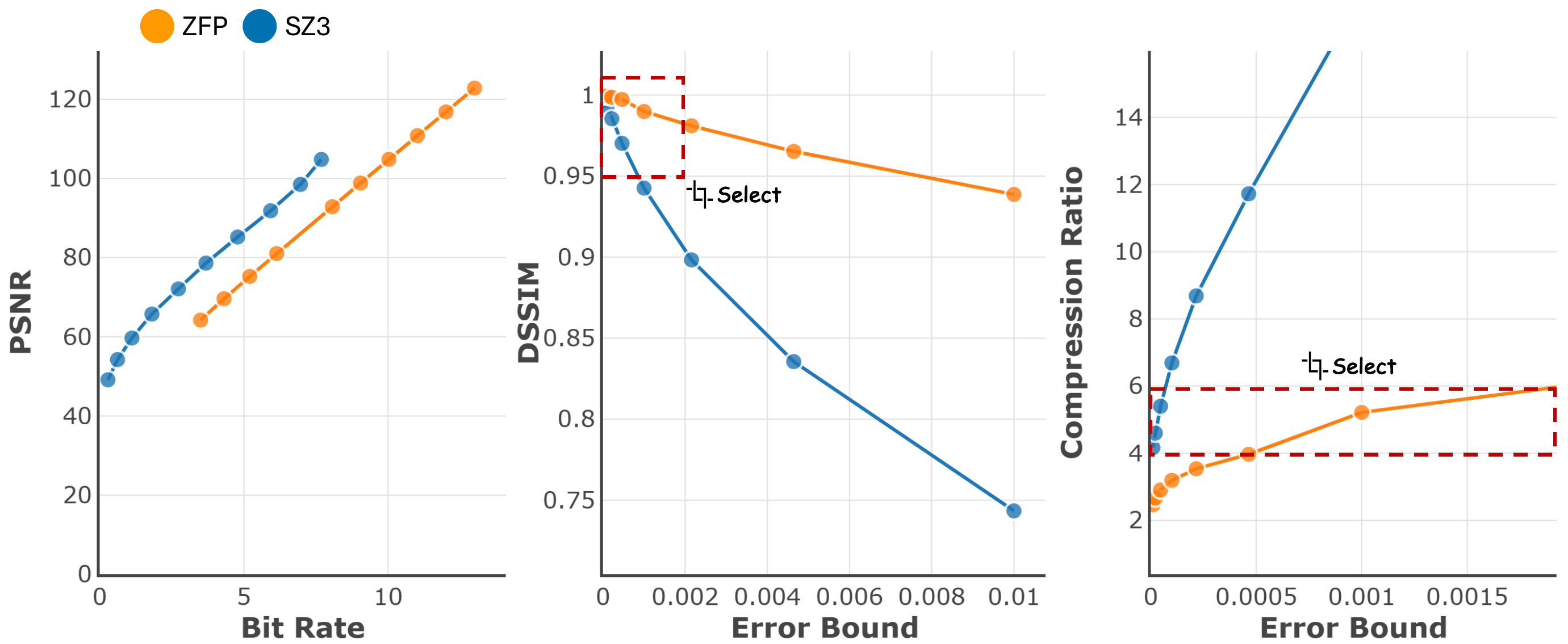}
    \caption{Performance comparison of SZ3 and ZFP on the CESM dataset across multiple error bound values. The figure summarizes the evaluated compressor instances using linked metric views, enabling users to compare trade-offs among bit rate, PSNR, DSSIM, and compression ratio.}
    \label{fig:cesm-evaluation}
\end{figure}

\paragraph{Evaluation Phase.} After this initial comparison, the user still does not know which error-bound range best balances quality and compression. FZ-VIS supports the next step by expanding the study across a broader range of error bounds (i.e., from $10^{-2}$ to $10^{-5}$) and aggregating the results in linked evaluation views. The user can switch among metric pairs, inspect where candidate points cluster, and use the ranked view to focus on promising regions of the design space. When PSNR is plotted against bit rate in \Cref{fig:cesm-evaluation}, the rate-distortion view confirms that SZ3 generally achieves lower bit rates, while ZFP tends to produce higher PSNR under similar error-bound values. The view thus supports targeted selection rather than ad hoc trial and error.

For climate applications, DSSIM and compression ratio are also important, so the workflow continues by switching the quantitative comparison from PSNR against bit rate to DSSIM against error bound. This interaction lets the user filter the candidates that satisfy DSSIM greater than $0.95$ and then compare compression ratios only among the admissible configurations. Under this requirement, SZ3 achieves a higher compression ratio than ZFP. For example, if the target compression ratio is around $4\times$, the user may choose SZ3 with a relative error bound of $2.15\times10^{-5}$, which yields a DSSIM of $0.99848$ and a compression ratio of $4.6$.

\subsection{Compressor Analysis for Developers}
Compressor developers need to understand the internal behavior of a compressor in order to improve its pipeline, rather than simply choose among existing options. In this case study, we show how FZ-VIS supports SZ3 developers by exposing intermediate outputs such as prediction residuals, quantization indices, and stage-wise runtime. These views help developers diagnose why a new module succeeds or fails and identify the next implementation steps.

\paragraph{Requirements.} Consider a realistic development scenario in which the SZ3 team evaluates an alternative regression-based predictor against the baseline Lorenzo predictor using the Hurricane Isabel dataset. To evaluate this new algorithm, developers need to determine whether the regression predictor improves residual structure, whether any such benefit survives quantization and entropy coding, and where it fails spatially. More specifically, they need tooling that reveals whether performance gains are confined to locally smooth regions, whether sharp transitions create large residual and quantization outliers, and whether the resulting compression-throughput trade-off justifies adoption or further optimization.

\paragraph{Design Phase.} Developers begin with the hypothesis that the regression-based predictor offers advantages, but they lack detailed insight into where it succeeds or fails within the pipeline. FZ-VIS supports this workflow by enabling them to evaluate the Lorenzo and regression-based predictors under identical settings, including error bound, quantizer, encoder, and lossless backend, while exposing intermediate artifacts for direct comparison. By inspecting residual and quantization-index distributions alongside stage-wise runtime breakdowns, developers can seamlessly trace behavior from prediction quality to downstream coding efficiency and runtime cost, rather than analyzing each metric in isolation.

\begin{figure}[htb]
    \centering
    \includegraphics[width=\linewidth, alt={Two histograms comparing Lorenzo and regression-based SZ3 predictors, with prediction residual distributions on the left and quantization-index distributions on the right.}]{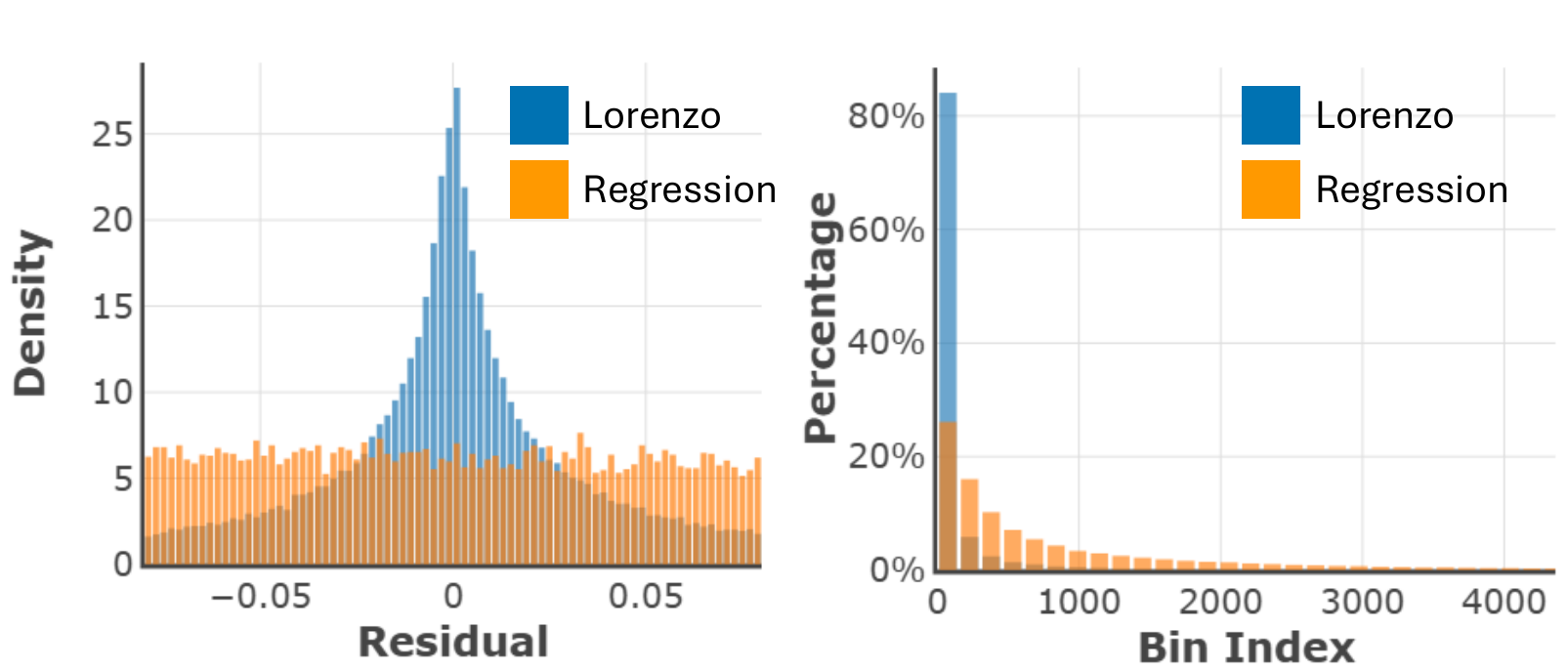}
    \caption{Comparison of the regression-based predictor and Lorenzo predictor in SZ3. Left: distribution of prediction residuals. Right: distribution of quantization indices.}
    \label{fig:residual-quantization-comparison}
\end{figure}

\begin{figure}[htb]
    \centering
    \includegraphics[width=\linewidth, alt={Stage-wise runtime comparison between Lorenzo and regression-based SZ3 predictors, with a stacked bar plot on the left and a table of additional compression metrics on the right.}]{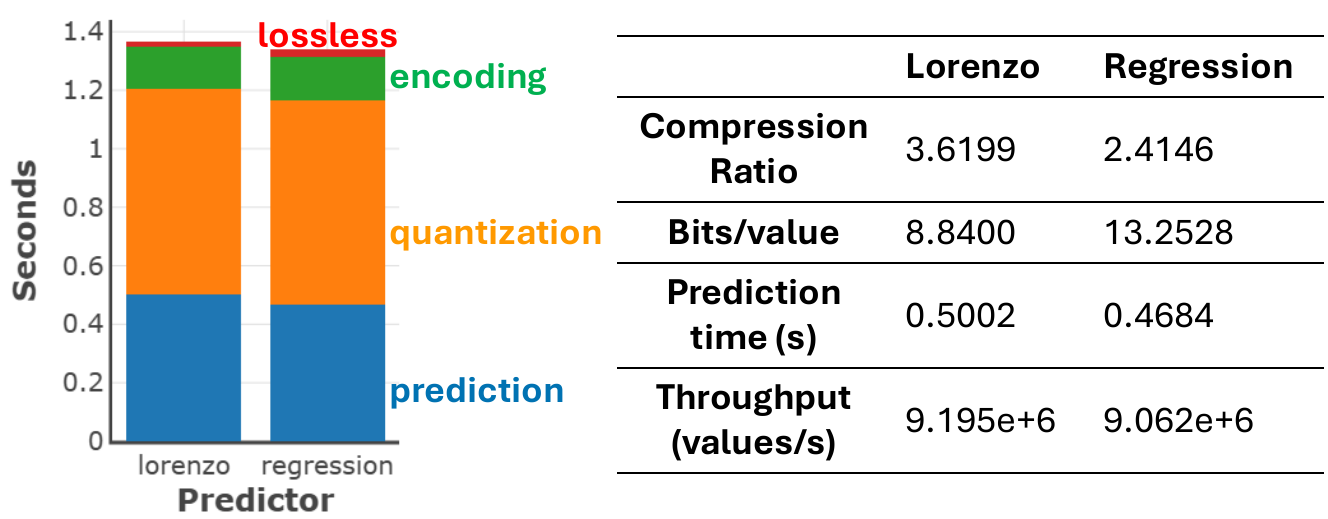}
    \caption{Stage-wise runtime comparison and additional metrics of the regression-based and Lorenzo predictors in SZ3 compressor.}
    \label{fig:stage-runtime}
\end{figure}

\paragraph{Evaluation Phase.} The linked views guide the developers through a staged diagnosis of the pipeline's performance. The residual histogram in \Cref{fig:residual-quantization-comparison} shows that while Lorenzo produces a noticeably higher concentration of near-zero residuals, the regression predictor exhibits a heavier residual tail. The quantization-index distributions then confirm that this weaker residual behavior persists downstream, translating directly into a wider index spread and poorer compression effectiveness. Finally, the stage-wise runtime breakdown in \Cref{fig:stage-runtime} demonstrates that the regression-based predictor does not offer a sufficient throughput advantage to offset its lower compression ratio. By moving from initial residual evidence to downstream coding behavior and final runtime costs, this workflow clearly illustrates why the baseline Lorenzo predictor remains the better choice for this particular dataset and pipeline configuration.

\subsection{Topological Feature Preservation}
Lossy compressors facilitate substantial data reduction while maintaining a predefined accuracy threshold. However, they may fail to preserve topological features, such as critical points, Morse-Smale (MS) complexes, and merge trees in decompressed data, even with bounded error~\cite{Liang2020Toward,Yan2024TopoSZ}. This case study shows how FZ-VIS guides domain scientists from diagnosing QoI failure to selecting a correction strategy. Specifically, we demonstrate how the framework enables users to choose a compressor that preserves specific topological features, including extrema and MS segmentations, based on their requirements.

\begin{figure}[htb]
    \centering
    \includegraphics[width=\linewidth, alt={Atmospheric river topology comparison with the original IVT field and incorrect Morse-Smale segmentation label counts across error bounds on the left, and examples of distorted segmentation boundaries for SZ3 and ZFP on the right.}]{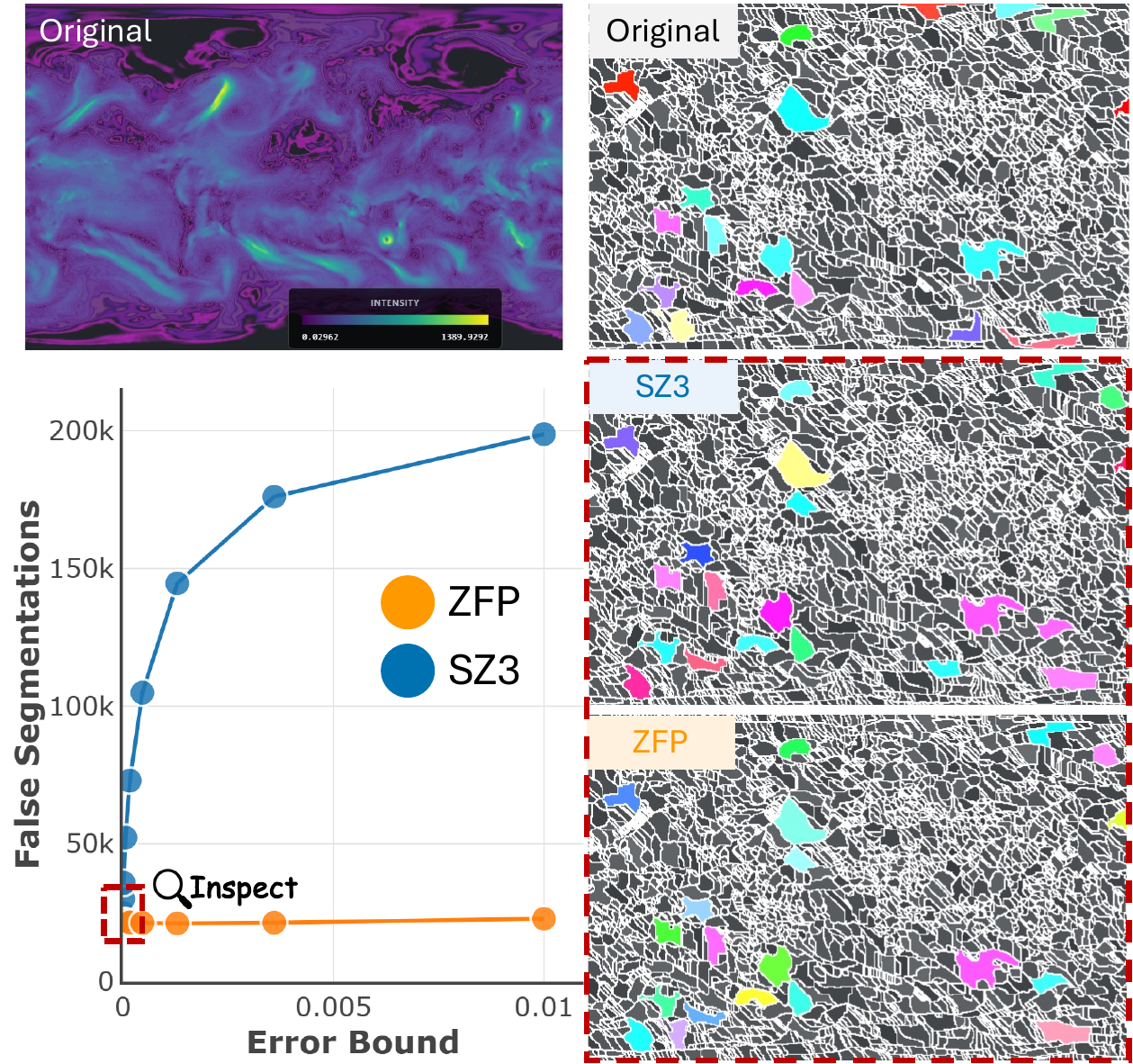}
    \caption{Impact of lossy compression on MS segmentations of the AR dataset. Left: visualization of the original dataset and the number of incorrect MS segmentation labels under varying error bounds for SZ3 and ZFP. Right: examples of distorted boundaries in the top 20 descending segmentations from decompressed SZ3 and ZFP data at a relative error bound of $10^{-5}$.}
    \label{fig:ar-ms-segmentation}
\end{figure}

\paragraph{Requirements.} In the analysis of Atmospheric Rivers (ARs), elongated bands of water vapor transport that originate from the tropics to North America and cause flooding, scientists characterize the skeleton of ARs with MS complexes and MS segmentations to better understand the formation and development of ARs~\cite{Lan2024Topological}. Distortions in such topological features could significantly impact scientists' understanding of AR formation and development, potentially leading to inaccurate evaluations of ARs' impact on precipitation and flooding in North America. Scientists characterize AR skeletons by the boundaries of descending segmentations of the Integrated Vapor Transport (IVT) field. Therefore, scientists need to choose a compressor that preserves MS segmentations for downstream data analysis tasks.
\paragraph{Design Phase.} At the outset, the scientist cannot predict whether simply tightening the error bound will recover the MS segmentation. FZ-VIS supports this diagnosis by coupling a quantitative QoI view with a linked spatial view. Through this interface, the user explores compression configurations across error bounds ranging from $10^{-2}$ to $10^{-5}$, computing critical points and MS segmentation on the decompressed data. As shown in \Cref{fig:ar-ms-segmentation}, ZFP generally preserves MS segmentations better than SZ3; however, even at the tightest tested error bound, more than 20,000 false segmentation labels remain. The linked spatial view explicitly reveals where distorted boundaries occur, which can affect the structural integrity of the AR features. Together, these views demonstrate that tightening error bounds alone is insufficient, thereby motivating a QoI-correction step.

\begin{figure}[htb]
    \centering
    \includegraphics[width=0.9\linewidth, alt={Line chart of overall compression ratio versus relative error bound after applying MSz correction to SZ3 and ZFP on the AR dataset.}]{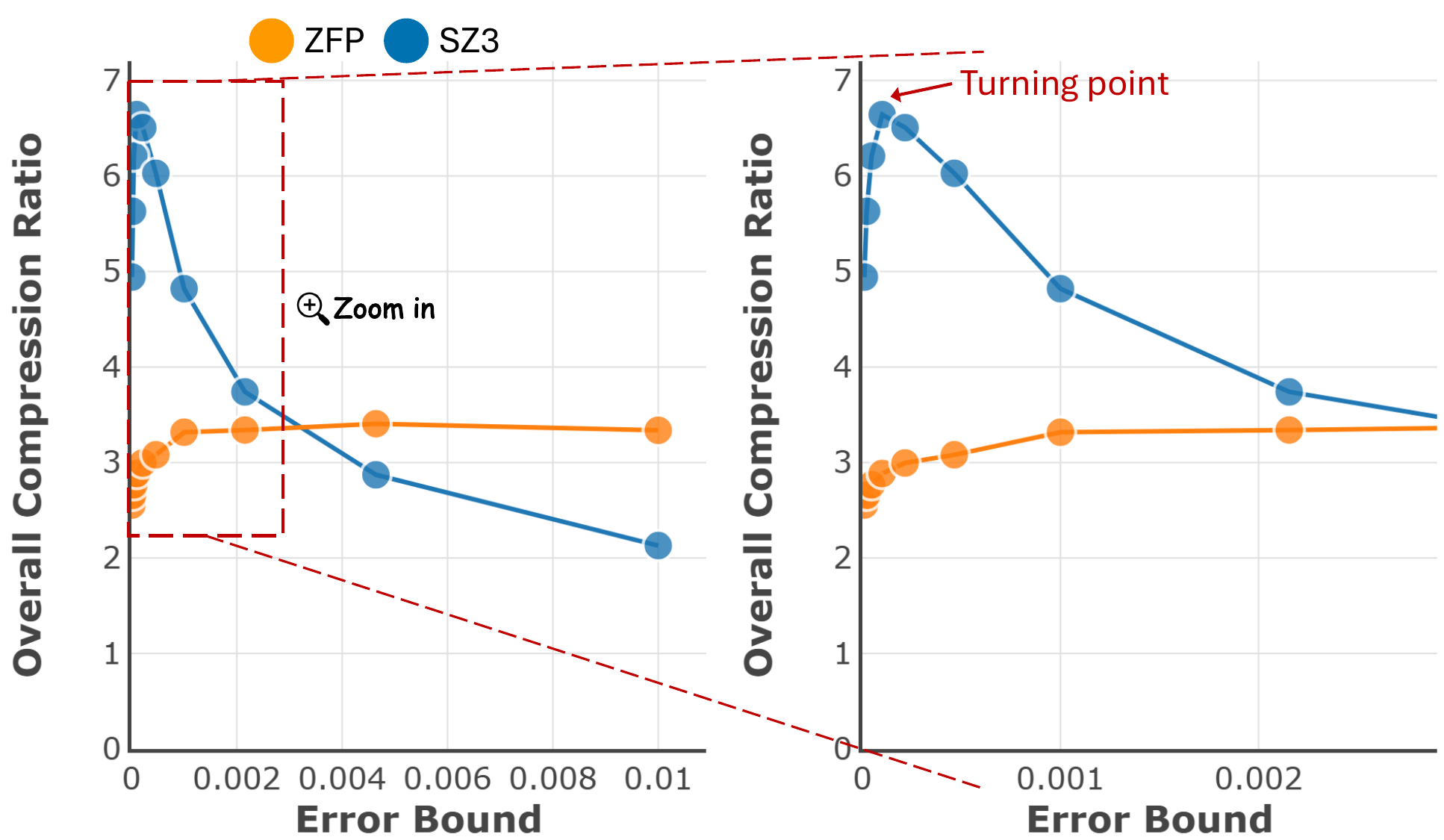}
    \caption{Overall compression ratio vs. error bound when applying MSz to preserve Morse-Smale segmentations of the AR dataset using SZ3 and ZFP compressors.}
    \label{fig:ocr-msz-ar}
\end{figure}

\paragraph{Evaluation Phase.} Once the initial diagnosis reveals that standard error-bounded compression alone is insufficient, FZ-VIS guides the user to the MSz correction module. The user evaluates the corrected candidates using the \textit{Overall Compression Ratio (OCR)} plot, where OCR is defined as the ratio of the original data size to the combined size of the compressed data and MSz edits~\cite{Li2025MSz}. This view exposes the trade-off between exact MS-segmentation recovery and compression efficiency. \Cref{fig:ocr-msz-ar} shows that SZ3 consistently attains a higher OCR than ZFP for error-bound values below $0.002$, with a favorable operating point near a relative error bound of $10^{-4}$.

\subsection{Spectral Feature Preservation}

While pointwise error control is important, the frequency-domain representation of data is also crucial for many scientific applications, as it provides complementary scientific insights. This case study shows how FZ-VIS extends a conventional spatial-error workflow with spectral evidence for compressor selection and refinement. We use the dark matter field from the NYX dataset to demonstrate this spectral workflow.

\paragraph{Requirements.} In cosmology applications, many analysis tasks rely on frequency-domain representations to capture key features. For example, the power spectrum of cosmology data is the key to analyzing matter and energy distribution across spatial scales. Therefore, scientists need to ensure that compressed data satisfies requirements in both the spatial and frequency domains. 

\begin{figure}[htb]
    \centering
    \includegraphics[width=\linewidth, alt={NYX spectral analysis showing the original data visualization, power spectrum and spectral relative error plots for ZFP configurations, and a table listing the error bounds and corresponding compression ratios.}]{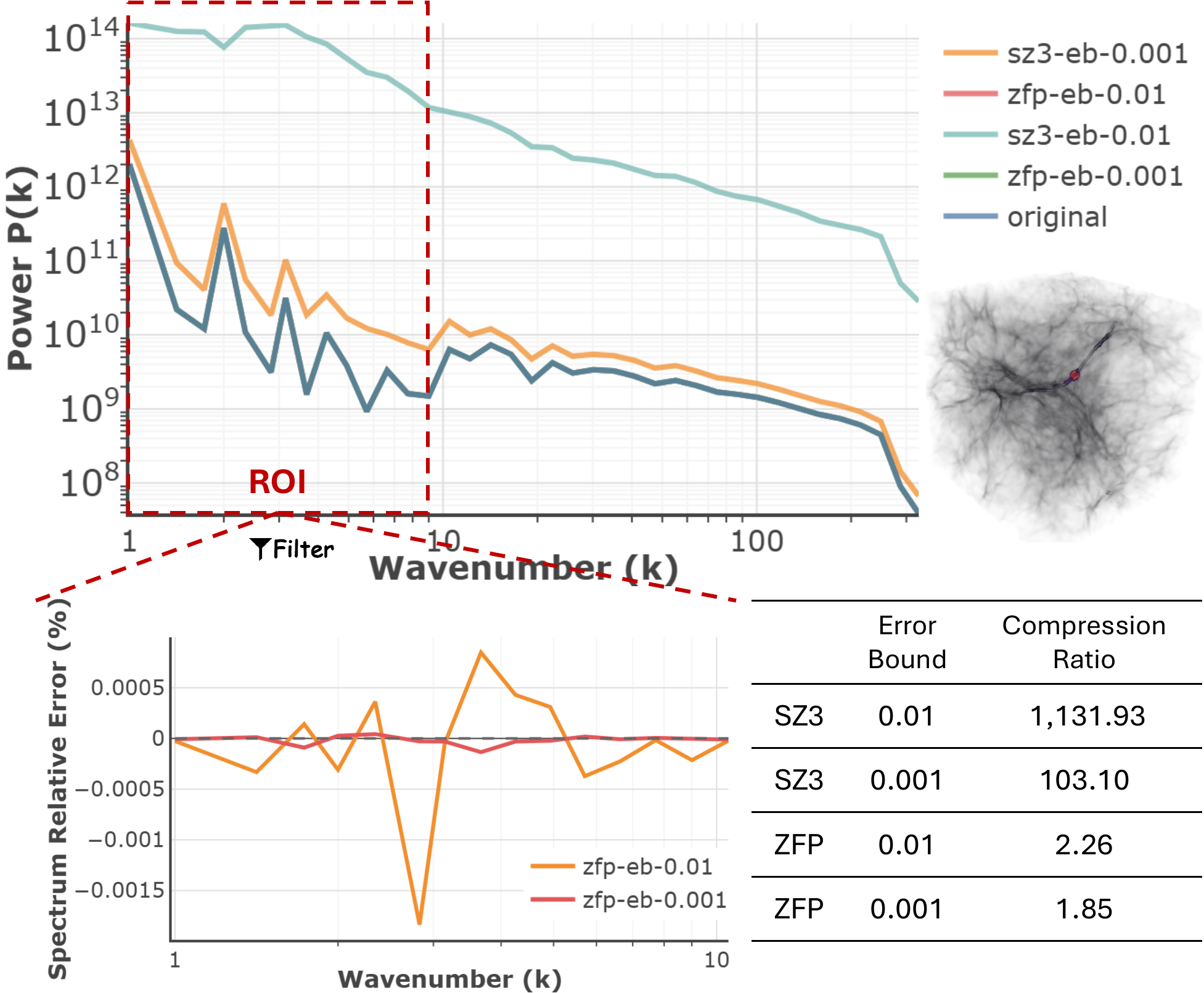}
    \caption{Top: power spectra of different compressor instances on the NYX dataset. Bottom: spectral relative error of ZFP in the selected frequency range, together with the corresponding compression ratios.}
    \label{fig:power-spectrum-comparison}
\end{figure}

\paragraph{Design Phase.} A scientist evaluating compressors solely through spatial error bounds may not know whether a configuration that satisfies pointwise thresholds also preserves the underlying data power spectrum. FZ-VIS bridges this gap by integrating dedicated spectral views into the workflow. When comparing SZ3 and ZFP at relative error bounds of $1\%$ and $0.1\%$, the user can simultaneously evaluate the power spectrum curves and their associated spectral relative errors. As shown in \Cref{fig:power-spectrum-comparison}, SZ3 introduces substantial spectral distortion at both tested error bounds, whereas ZFP better preserves the trend of the original power spectrum and keeps the spectral relative error below $0.0020\%$ within the selected frequency ROI. However, this improved spectral fidelity comes at the cost of compression performance: at a $1\%$ relative error bound, ZFP achieves a compression ratio of only $2.26$, compared to SZ3's significantly higher ratio of $103.10$.

\begin{figure}[htb]
    \centering
    \includegraphics[width=\linewidth, alt={FFCz correction module with a user interface for sweeping the frequency error bound on the left and a plot of SZ3 compression ratio versus frequency error bound at fixed spatial error on the right.}]{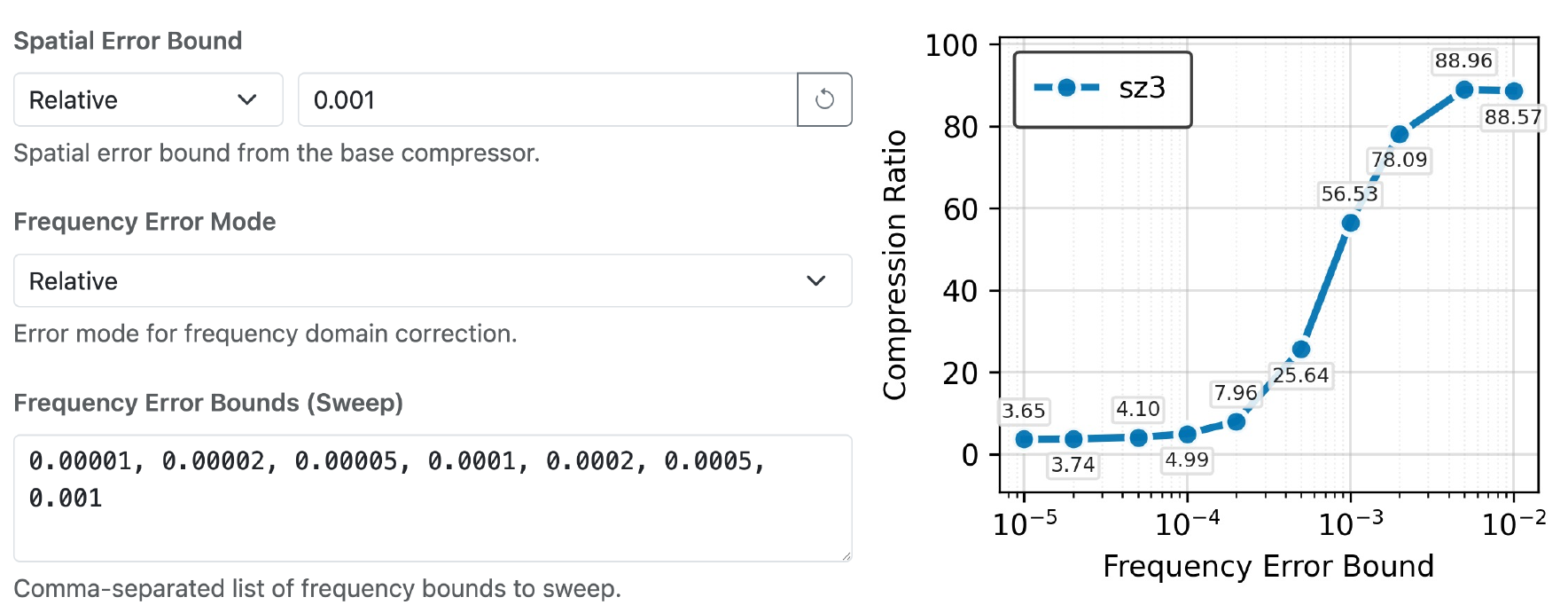}
    \caption{Left: the FFCz correction module allows users to sweep the frequency error bound. Right: compression ratio against frequency error bound for SZ3 with a spatial error bound of $0.1\%$ on the NYX dataset.}
    \label{fig:ffcz-sz3}
\end{figure}

\paragraph{Evaluation Phase.} Recognizing that direct compressor selection forces a difficult trade-off between spectral fidelity and data reduction, the user next turns to refinement rather than abandoning the high-ratio SZ3 candidate. FZ-VIS supports this step by integrating FFCz~\cite{Ren2026FFCz} and visualizing the trade-off between compression ratio and spectral error bound under a fixed spatial tolerance. \Cref{fig:ffcz-sz3} lets the user sweep the frequency-domain constraint and evaluate how aggressively it can be tightened before the compression ratio drops unacceptably. When the user sets the frequency error bound to $0.1\%$ with a spatial error bound of $0.1\%$, the resulting compression ratio is still around $56$, far higher than directly choosing ZFP.

\section{Limitations and Discussion}
\label{sec:limitations}

FZ-VIS is designed to make QoI-aware lossy compression exploration more accessible. However, it does not eliminate all practical challenges in compressor design and evaluation. 

First, its current composition model primarily supports existing compressors and predefined pipelines exposed through LibPressio. Although users can compare module options and parameter variants, they cannot yet freely construct a new lossy compressor by arbitrarily composing modules. Enabling this capability would require careful consideration of module compatibility and execution constraints. 

Second, the current QoI support in FZ-VIS focuses on specific features, such as critical points, Morse-Smale segmentation, and power spectrum, whereas many applications require additional task-specific quantities and derived measures. 

Finally, the current web-based interactive framework may not scale well to very large datasets because data movement, rendering, and repeated analysis can degrade responsiveness during exploration.

\section{Conclusion}
\label{sec:conclusion}

In this paper, we presented FZ-VIS, a human-in-the-loop visual analytics framework for QoI-aware lossy compression. FZ-VIS unifies compressor configuration, batch evaluation, and QoI analysis within a single workflow, enabling users to explore compressor variants, examine trade-offs between compression performance and data fidelity, and evaluate feature preservation for downstream analysis. By integrating an intuitive web-based interface, configuration generation via a force-directed graph, linked visual and quantitative analysis views, and QoI-oriented modules for topological and spectral analysis, FZ-VIS supports novice users, compressor developers, and domain scientists within a unified environment.

More broadly, FZ-VIS connects the compressor design space to the scientific evaluation space, allowing users to reason jointly about configuration choices, conventional fidelity metrics, and application-specific QoIs during the decision-making process. Our case studies show that FZ-VIS helps users navigate complex design spaces and make informed compression decisions across diverse scientific tasks.

We view FZ-VIS as a foundation for more flexible and scalable QoI-aware compression analysis, and as a step toward making compression design more accessible, interpretable, and scientifically grounded across diverse applications. Looking forward, possible directions include supporting more flexible compressor composition, broadening QoI coverage to additional scientific tasks, and improving scalability for larger datasets and more demanding interactive analysis tasks.

\section*{Supplemental Materials}
\label{sec:supplemental_materials}
All supplemental materials are available on OSF at \url{https://osf.io/7cspq}, released under a CC BY 4.0 license. They include (1) a video demonstration of the proposed system, (2) a formative user study for requirements analysis, and (3) an extended version of this paper that includes the appendix. The source code for the implementation is also available at the same link under the MIT license.

\section*{Acknowledgment}
The material was supported by the U.S. Department of Energy, Office of Science, Advanced Scientific Computing Research (ASCR), under contract DE-AC02-06CH11357, and supported by the National Science Foundation under grants OAC-2311875, OAC-2311878, OAC-2514036, OAC-2514034, OAC-2609480, OAC-2513768, OAC-2313122, OAC-2313123, OAC-2313124, and OAC-2628473. 

\bibliographystyle{abbrv-doi-hyperref}
\bibliography{refs}

\newpage
\appendix 
\crefalias{section}{appendix}
\section{Workflow Comparison}
\label{sec:appendix-workflow-comparison}

\noindent This section summarizes how FZ-VIS differs from existing tools in compression workflow and QoI-aware analysis capabilities. \Cref{tab:workflow-comparison} complements the discussion in \Cref{sec:related} and \Cref{sec:system-design}. While LibPressio~\cite{Underwood2021Productive} offers a unified interface for configuring compressors, ParaView~\cite{Ayachit2015ParaView} provides strong post hoc data visualization, and tools like Z-checker~\cite{Tao2019Z-checker} or Foresight~\cite{Grosset2020Foresight} support comparative compression analysis, these capabilities are typically distributed across separate environments. In contrast, FZ-VIS combines interactive composition, batch generation of configuration variants, linked quantitative and spatial inspection, intermediate-stage analysis, and QoI-aware refinement within a unified workflow. In the table, \emph{Partial} indicates that a tool supports only part of the functionality or requires substantial manual work, whereas \emph{Limited} indicates support restricted to narrow cases rather than general QoI-aware analysis.

\begin{table}[htb]
    \centering
    \scriptsize
    \caption{Workflow-level comparison of FZ-VIS and existing tools: LP (LibPressio), PV (ParaView), and ZC / FS (Z-checker / Foresight). The comparison focuses on integrated support for QoI-aware compression design rather than on compression quality itself.}
    \label{tab:workflow-comparison}

    \renewcommand{\arraystretch}{1.3}
    \begin{tabular}{@{}>{\raggedright\arraybackslash}m{0.42\linewidth}cccc@{}}
        \toprule
        \textbf{Capability} & \textbf{LP} & \textbf{PV} & \textbf{ZC / FS} & \textbf{FZ-VIS} \\
        \midrule
        Interactive compressor composition & N & N & N & Y \\
        \hline
        Batch generation of configurations & Y & N & Lim. & Y \\
        \hline
        Compression metrics comparison & Y & Part. & Y & Y \\
        \hline
        Linked quantitative and spatial comparison & N & Part. & Part. & Y \\
        \hline
        Inspection of intermediate compression outputs & Y & N & Lim. & Y \\
        \hline
        Integrated QoI-specific analysis & Part. & Part. & N & Y \\
        \hline
        QoI-guided refinement / correction & Lim. & N & N & Y \\
        \hline
        Single-environment workflow (no external scripting) & N & N & N & Y \\
        \bottomrule
    \end{tabular}
\end{table}

The rows of \Cref{tab:workflow-comparison} reflect the main stages of a QoI-aware compression workflow. The first two rows cover design-space construction: whether users can interactively compose compressor pipelines and systematically generate configuration families. The middle rows emphasize comparative analysis, covering whether quantitative summaries can be linked with spatial inspection and whether intermediate pipeline outputs can be examined to diagnose compressor behavior. The QoI-related rows capture whether the environment supports application-specific validation and correction beyond conventional fidelity metrics.

The comparison highlights that FZ-VIS distinguishes itself through workflow integration rather than replacing existing tools. LibPressio remains valuable as a modular compression interface, ParaView remains strong for scientific visualization, and Z-checker and Foresight remain useful for benchmarking and error analysis. However, when scientists need to iteratively move between compressor configuration, result inspection, QoI validation, and refinement, existing approaches often require external scripts or switching between tools. The last row captures a key practical advantage: FZ-VIS reduces coordination overhead by consolidating these steps into one environment, which is particularly valuable for exploratory, human-in-the-loop QoI analysis.

\end{document}